\documentclass[aps,pra,reprint,amsmath,amssymb,showpacs,groupedaddress,floatfix]{revtex4-1}
\usepackage{mathtools,graphicx,bbold,bm,color,url}
\usepackage[colorlinks,citecolor=blue,urlcolor=blue,linkcolor=blue,hyperindex,breaklinks]{hyperref}

\newcommand{\dd}{\text{d}}
\newcommand{\la}{\langle}
\newcommand{\ra}{\rangle}
\newcommand{\lla}{\left\langle}
\newcommand{\rra}{\right\rangle}
\newcommand{\lam}{\lambda}
\newcommand{\mcP}{\mathcal{P}}

\newcommand{\One}{\mathbb{1}}
\newcommand{\tsum}{\textstyle\sum}	% careful, these need bracketing
\newcommand{\tprod}{\textstyle\prod}	% careful, these need bracketing
\graphicspath{{./figures/}}
\begin{document}

\title{Eigenvalue Determination for Mixed Quantum States\\ 
   using Overlap Statistics}

\author{L\'azaro Alonso}
\email{lazarus.alon@gmail.com}
\affiliation{Instituto de F\'isica, Benem\'erita Universidad Aut\'onoma de Puebla, Puebla, M\'exico}

\author{David Bermudez}
\email{dbermudez@fis.cinvestav.mx}
%\homepage{\\ http://www.fis.cinvestav.mx/{$\sim$}dbermudez/}
\affiliation{Departmento de F\'isica, Cinvestav, A.P. 14-740, 07000 Ciudad de M\'exico, M\'exico}

\author{Thomas Gorin}
\email{thomas.gorin@cucei.udg.mx}
\affiliation{Departamento de F\'isica, Universidad de Guadalajara,  Guadalajara, Jalisco, M\' exico}
\date{\today}

\begin{abstract}
We consider the statistics of overlaps between a mixed state and its image 
under random unitary transformations. Choosing the transformations from the
unitary group with its invariant (Haar) measure, the distribution of overlaps
depends only on the eigenvalues of the mixed state. This allows one to estimate 
these eigenvalues from the overlap statistics. In the first part of this work, 
we present explicit results for
qutrits, including a discussion of the expected uncertainties in the eigenvalue
estimation. In the second part, we assume that the set of
available unitary transformations is restricted to $SO(3)$, realized as Wigner
$D$-matrices. In that case, the overlap statistics does not depend only on the
eigenvalues, but also on the eigenstates of the mixed state under scrutiny. The
overlap distribution then shows a complicated pattern, which may be considered
as a fingerprint of the mixed state. When using random transformations from the unitary group, the eigenvalues can
be determined quite simply from the lower and the upper limit of the overlap 
statistics. This may still be possible in the $SO(3)$ case, but only at the 
expense of a finite systematic uncertainty. \\

%\noindent
PhySH: quantum parameter estimation, quantum tomography, group theory, quantum measurements.

\end{abstract}


\maketitle

\section{\label{I} Introduction}

Quantum state tomography~\cite{ParisRehacek04} deals with the complete
characterization of an ensemble of identically and independently distributed
quantum states. It involves the estimation of all elements of the density 
matrix describing that ensemble of states. As a consequence, experimental 
protocols become expensive and time consuming, even for medium sized quantum 
systems~\cite{Hae05,Cra2010}. A natural reduction of the tomography problem 
consists of only estimating the eigenvalues of the density matrix. For such an 
approach, different experimental schemes have been 
proposed~\cite{KeyWer01,Bal06,Tan14,EnkBee12}.

The method to be discussed here is based on measuring the overlap between two
mixed quantum states~\cite{Walborn2006,Walborn2007}. This requires two 
copies of the quantum state, the possibility to apply unitary transformations 
to one of these copies, and finally a measurement of the overlap between the 
two states. Different schemes for overlap measurements are discussed for 
instance in Refs.~\cite{Filip02,BaLeMi13}, while the experimental application 
of unitary transformations (together with complete tomography) has been shown, 
e.g., in ``transmon'' qutrits~\cite{Bianchetti10}; a variant of the 
superconducting charge qubit. In Ref.~\cite{Vitanov12}, several scenarios have 
been proposed, which allow to realize general $SU(3)$ transformations in 
atomic three-level systems.

Our method is based on the fact that the distribution of overlaps 
$P_\varrho(q)$ between a given density matrix $\varrho$ and its image under 
random unitary transformations depends only on the eigenvalues of $\varrho$. 
These eigenvalues may then be estimated from the knowledge of $P_\varrho(q)$. 
For that, it is crucial to compute $P_\varrho(q)$ analytically, as this allows 
us to find convenient strategies to solve the inverse problem of estimating the 
eigenvalues. In order to compute $P_\varrho(q)$, we generalize previous 
results on the joint probability distribution of projection probabilities for 
random orthonormal quantum states~\cite{AG16}. We then concentrate on the 
qutrit case, where we obtain $P_\varrho(q)$ in closed form.

There may be experimental situations, where the sampling over all 
unitary transformations is not possible, i.e., $SU(3)$ in the qutrit case.
Therefore, we also consider the distribution of overlaps, when the 
transformations are restricted to rotations in real space. This amounts to
consider a sub-group of $SU(3)$ that can be parametrized by the respective
Wigner $D$-matrices~\cite{Sakurai94}. In this case, the overlap distribution
actually contains more information about the original density matrix $\varrho$,
not only the eigenvalues. However, in the absence of an analytical expression, 
the extraction of that information is much more difficult. Nevertheless, even 
in that case, our approach can provide important bounds for the eigenvalues
of $\varrho$.

The paper is organized as follows: After this introduction, 
in Sec.~\ref{G},   %General definitions and conventions
we provide the mathematical and technical background for the  overlap function.
In Sec.~\ref{M},   %JPD for the general case
we present the analytical calculation of the overlap distribution for the 
$SU(3)$ case.
In Sec.~\ref{S},   %Stability of the eigenvalue estimation
we compute the expected uncertainties of the eigenvalue estimation 
based on the upper and lower limit of the overlap distribution, themselves
having been estimated with finite precision.
In Sec.~\ref{W},    %Overlap function and Wigner $D$-matrices
we discuss overlap distributions for transformations limited to $SO(3)$.
Conclusions are provided in Sec. ~\ref{refC}.

\section{\label{G} General definitions and notation}

Let $\mathcal{H}$ be the Hilbert space of finite dimension 
$N={\rm dim}(\mathcal{H})$ corresponding to some quantum system. Let $\varrho$ 
be a mixed state with $N$ non-negative eigenvalues 
$\bm{\lam}= (\lam_1,\lam_2,\ldots,\lam_N)$, normalized as 
$\sum_{j=1}^N \lam_j = 1$. 
We are then interested in the distribution of the values of the overlap 
function
\begin{equation}
Q(O,\varrho)= {\rm tr}\big (\, O\, \varrho\, O^\dagger\, \varrho\, \big ) ,
\end{equation}
averaged over the whole unitary group with the respective Haar 
measure~\cite{Haar33,Weyl39}. In a different context, the same quantity has
been considered as a measure for the distance between the reduced dynamics of
two open quantum systems~\cite{ZP03,GPSZ06}. Note that $Q(O,\varrho)$ does not 
depend on the value of the determinant of $O$, which means that the 
distribution over the unitary group or the corresponding special unitary group 
yields exactly the same result. 
Note also that for $O = \One$ the overlap function reduces to the purity, used 
in the area of quantum information~\cite{NieChu00}.

%Here, we will restrict ourselves to the unitary group. 
Our main interest is in the probability density 
\begin{equation}
P_{\varrho}(q) = \langle \delta[q - Q(O,\varrho)]\rangle
 = \langle \delta[q - Q(O,\bm{\lam})]\rangle\; ,
\end{equation}
where the angular brackets represent the average over the unitary group with 
respect to the normalized Haar measure. The second equality is valid due to the 
invariance of this measure. It means that $P_{\varrho}(q)$ depends only on the 
eigenvalues of $\varrho$ ~%
\footnote{The shape of $\bm{\lam}$ may be varying between vector and diagonal
  matrix form, depending on the context.}.
In what follows, we will assume that the eigenvalues are arranged in ascending 
order:
$\lam_{\rm min} = \lam_1 \le \lam_2 \le \ldots \le \lam_N = \lam_{\rm max}$. 
This can always be achieved, since permutations belong to the unitary group. 
Now, we may write,
\begin{align}
&Q(O,\bm{\lam})= \sum_{j,k=1}^N \lam_j \lam_k |O_{jk}|^2 
   = \sum_{k=1}^N t_k \lam_k \; , \notag\\ 
&\text{where}\quad t_k = \sum_{j=1}^N \lam_j |O_{jk}|^2 \; .
\label{overlapFunc}\end{align}
The normalization of the column vectors of $O$ implies that
$|O_{jN}|^2 = 1- \sum_{k = 1}^{N-1} |O_{jk}|^2$. Using this, and the fact that 
density matrices have unit trace $\sum_{j=1}^N \lam_j = 1$, we may reduce the
number of partial overlaps by one and write
\begin{align}
Q(O, \bm{\lam})
&= \lam_{N} + \sum_{k = 1}^{N-1}t_{k}\, (\lam_{k} - \lam_{N}) \; .
\label{G:QforNm1}\end{align}

Therefore, we can calculate the overlap distribution $P_\varrho(q)$ from the 
joint probability distribution (JPD) as
\begin{equation}\label{pnklambda}
\mcP_{\bm{\lam}}(\bm{t}') = \lla {\tprod_{k=1}^{N-1} 
   \delta\left[ t_k - \tsum_{j=1}^N \lam_j |O_{jk}|^2\right] }\rra ,
%\label{I:PNKdef1}
\end{equation}
for the partial overlaps $\bm{t}' = (t_1, t_2, \ldots, t_{N-1})$, by the 
following $N-1$ dimensional integral:
\begin{equation}
P_\varrho(q) = \left\{\prod_{k=1}^{N-1} \int_0^1{\rm d} t_k \right\}
   \mathcal{P}_{\bm{\lam}}(\bm{t}')\; 
   \delta\big [ q - Q(O, \bm{\lam})\, \big ] .
\end{equation}
The JPD $\mcP_{\bm{\lam}}(\bm{t}')$ can be calculated by generalizing an earlier 
work on the projection probabilities of random orthonormal states~\cite{AG16}.

In the following  Sec~\ref{G2}, we discuss some special cases, where the
calculation of the overlap distribution is particularly simple. This may help
to prepare the ground for working out the general case, considered in 
Secs.~\ref{M} and~\ref{S}.

\subsection{\label{G2} Special cases for \boldmath $N = 2$ and $N = 3$}

In the qubit case, $N = 2$, the overlap distribution can be obtained in 
general. In the qutrit case, $N=3$, we will consider special cases, where one
or two eigenvalues of the density matrix are equal to zero.

In the qubit case, we find from Eqs.~(\ref{overlapFunc}) and~(\ref{G:QforNm1}) 
that
\begin{align}
&Q(O, \varrho) = \lam_2 + (\lam_1 - \lam_2)\; t_1 \; , \\
&t_1 = \lam_1 |O_{11}|^2 + \lam_2 |O_{21}|^2
 = \lam_2 + (\lam_1 - \lam_2)\; |O_{11}|^2 \; ,
\notag\end{align}
such that the overlap function is given by
\begin{equation}\label{qubit}
Q(O,\varrho) =  2\lam_1\lam_2 +  (\lam_1 - \lam_2)^2\; |O_{11}|^2 \; .
\end{equation}
The overlap depends linearly on the absolute value squared of the random 
unitary matrix $O\in U(2)$. In Ref.~\cite{AG16}, the distribution of sums of $K$
such absolute values squared have been calculated as averages over $U(N)$,
denoted by $\mathcal{P}_{NK}(t)$. For the present case, we may set 
$t= |O_{11}|^2$ and find $\mathcal{P}_{21}(t)= \Theta(t)\, \Theta(1-t)$, where 
$\Theta(t)$ is the unit step function.

\begin{figure} 
\includegraphics[scale= .43]{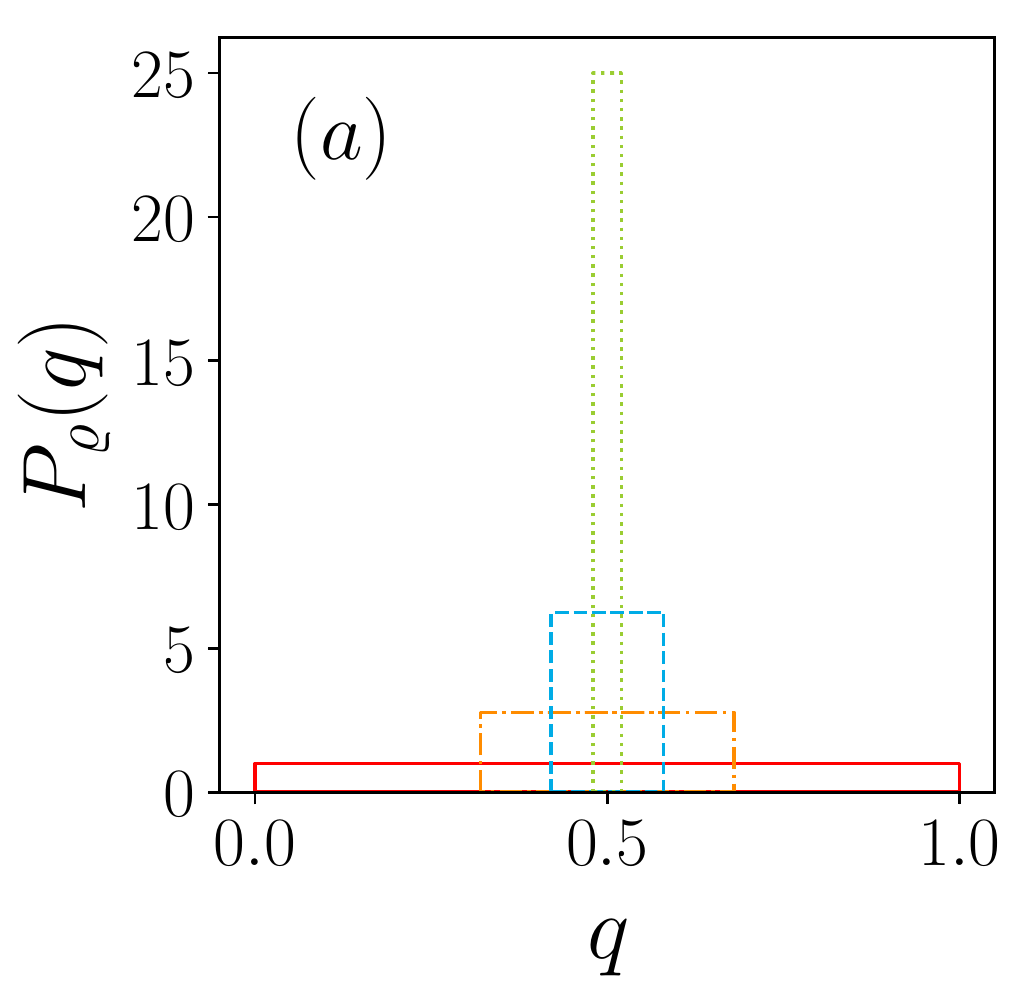}
\includegraphics[scale= .43]{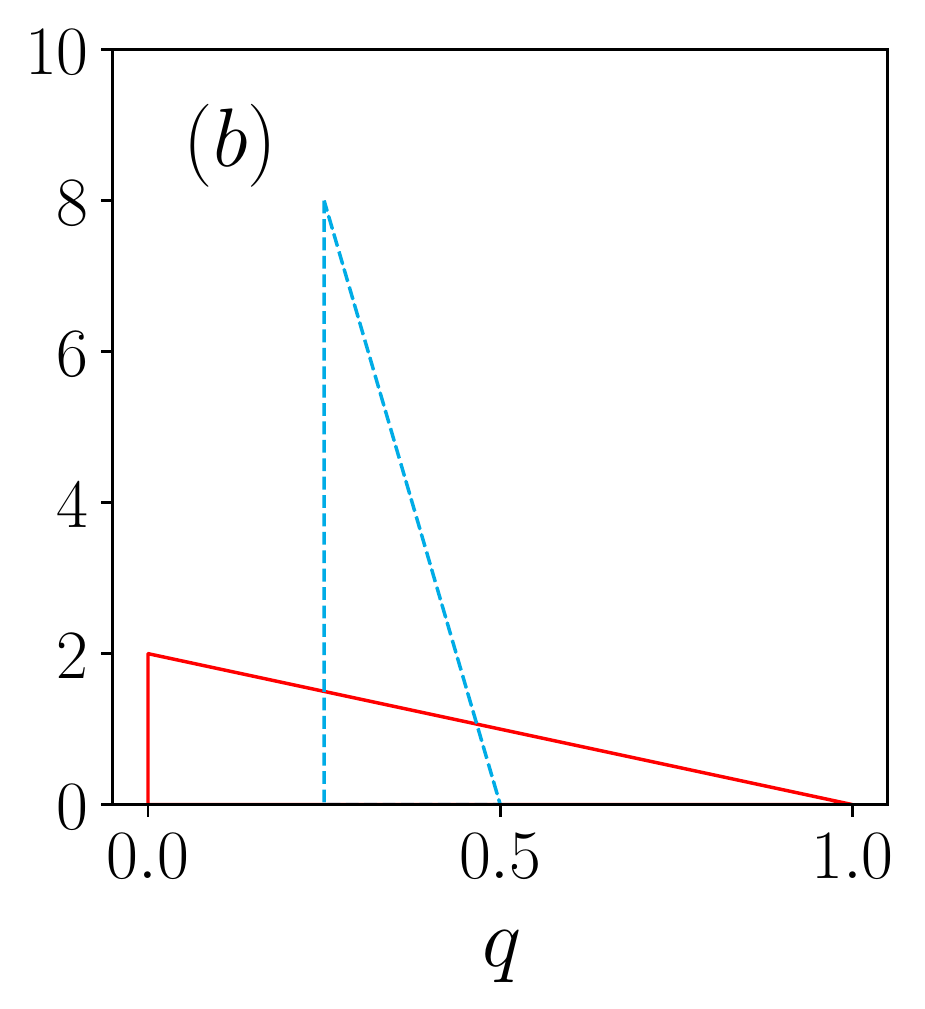}
\caption{(Color online) Overlap probability density $P_{\varrho}(q)$ for 
$N= 2\, (3)$ and different sets of eigenvalues. Panel (a) shows the qubit 
case: $\lam_1\, (\lam_2) =0\, (1)$ (red solid line), 
  $\lam_1\, (\lam_2) = 0.2\, (0.8)$ (orange dot dashed line), 
  $\lam_1\, (\lam_2) = 0.3\, (0.7)$ (blue dashed line), and
  $\lam_1\, (\lam_2) = 0.4,  (0.6)$ (green dotted line). Panel (b) shows the 
qutrit case:  $\lam_1 =\lam_2 = 0\, ,\; \lam_3 = 1$ (red solid line), and 
  $\lam_1 = 0\, ,\; \lam_2 = \lam_3 = 0.5$ (blue dashed line).}
\label{distQubit}\end{figure}

With the maximum (minimum) value for $Q(O,\varrho)$ is given by
\begin{equation}
q_{\rm max} = \lam_1^2 + \lam_2^2\quad\text{and}\quad
q_{\rm min} =  2\lam_1\lam_2 \; ,
\end{equation}
respectively, we obtain
\begin{align}
P_{\varrho}(q) &= \la\, \delta\left[ q - Q(O,\varrho)\right] \,\ra \notag\\
&= \int_{0}^{1} {\rm {d}}t \;\mathcal{P}_{21}(t)\;  
   \delta\big [\, q - (2\lam_1\lam_2 +  (\lam_1 - \lam_2)^2\; t)\, \big] 
   \notag\\
&= \frac{\Theta(q - q_{\rm min})\, \Theta(q_{\rm max} - q)}
   {(\lam_1 - \lam_2)^2} \; .
\end{align}
In Fig.~\ref{distQubit} (a), we show the distribution $P_\varrho(q)$ for 
several values of  $\lam_1$ and $\lam_2$. A delta distribution is reached when 
$\lam_1 = \lam_2$. Due to  the unit trace property, $\lam_1 + \lam_2 = 1$, any 
single quantitative property of the overlap distribution is enough to estimate 
the eigenvalues of $\varrho$. This quantity might be its maximum  value, 
$(\lam_1 - \lam_2)^{-2}$, $q_{\rm max}$ or $q_{\rm min}$. Experimentally, the 
easiest case would consist in measuring $q_{\rm max}$ which turns out to be the 
purity of our the mixed state, showing that a random sampling can be spared 
entirely in this case.

Even for the mixed state of a qutrit ($N=3$), it is possible to obtain simple
expressions for the overlap distribution, in some special cases. The line of
argument is analogous to the qubit case, and we delegate the corresponding 
discussion to App.~\ref{G3}. In Fig.~\ref{distQubit}(b), we show two of these 
cases, $\lam_1= 0\, , \; \lam_2= \lam_3= 1/2$, where
\begin{equation}
P_{\varrho}(q)=  16\, (1 - 2q)\; \Theta(q - 1/4)\; \Theta(1/2 - q)\; ,
\end{equation}
and $\lam_1= \lam_2= 0\, , \; \lam_3= 1$, where
\begin{equation}
P_{\varrho}(q) = 2\, (1 - q)\; \Theta(q)\; \Theta(1 - q)\; .
\end{equation}
In order to estimate the eigenvalues of an arbitrary mixed qutrit 
state, we would need two quantitative properties of the overlap distribution.
In this work we will concentrate on the limits $q_{\rm max}$ and $q_{\rm min}$
 of $P_\varrho(q)$.

\section{\label{M} JPD for general case}

In this section, we derive a general integral expression for the joint 
probability density defined in Eq.~(\ref{pnklambda}). We will closely follow 
the derivation for the unitary case in Ref.~\cite{AG16}. According to this
work, Eq.~(\ref{pnklambda}) can be written as 
\begin{align}
 &\mcP_{\bm\lam}(\bm{t}') \propto \left\{ {\tprod_{k=1}^{N-1}} 
\int d\Omega_2(\bm{w}_k)\;\right.\notag\\
&\times \left.\delta\Big ( t_k - {\tsum_{j=1}^N}\lambda_j |w_{jk}|^2 \Big )\right\}\; 
\prod_{\mu < \nu}^{N-1} 
\delta^2\big ( \la\bm{w}_\mu\, |\, \bm{w}_\nu\ra \big ) \; .
\label{VRU:PNKdef}
\end{align}
Here, $d \Omega_2(\bm{w}_\xi)$ denotes the uniform measure on the hypersphere 
in $\mathbb{R}^{2N}$, which implements the normalization of the column vectors 
$\bm{w}_\xi$ of $O$. Correspondingly, the last product of delta functions 
implements the orthogonality conditions between these column vectors. In that
case, the delta-function is two-dimensional because for 
$\la\bm{w}_\mu | \bm{w}_\nu\ra$ to be zero, the real and the imaginary part 
must be zero. The integral expression on the RHS is proportional to 
$\mcP_{\bm\lam}(\bm{t}')$ such that one has to compute the normalization 
constant for $\mcP_{\bm\lam}(\bm{t}')$, separately.

Following the steps described in Ref. \cite{AG16}, we arrive at the following
result:
\begin{align}
\mathcal{P}_{\bm{\lam}}({\bf{t}}') &=  \frac{1}{Z(\bm{o})}
   \left\{\prod_{k = 1}^{N -1}\int\frac{ds_{k}}{2\pi}\right\} Z(\bm{s})\; ,
\end{align}
where $Z(\bm{s})$ and the matrix $\bm{C}$, it depends upon, are defined as
\begin{align}
Z(\bm{s}) &= 
  \left\{\prod_{\mu<\nu}^{N-1}\int\frac{d^2\tau_{\mu\nu}}{4\pi^2}\right\} 
  \frac{1}{\prod_{j = 1}^{N} \det[\bm{C}(\lam_j)]}\; ,\label{zetas}\\
\bm{C}(\lam) &= \left(\begin{array}{ccc} 
c_1(\lam) & \tau/2& \ldots \\
-\tau^*/2 & c_2(\lam) &\\
\vdots & & \ddots
\end{array}\right)\; ,
\end{align}
with $c_j(\lam)= 1 - is_j (\lam - t_j)$. The matrix $\bm{C}$ is a $N-1$ 
dimensional square matrix which contains the  $\mathbf{t'}$  parameters. In
the following Sec.~\ref{M3}, we compute $\mathcal{P}_{\bm{\lam}}({\bf{t}}')$
for the qutrit case $N=3$.

\subsection{\label{M3} JPD for \boldmath $N=3$}

Let us assume the ordering $\lam_1 < \lam_2 < \lam_3$, and write the overlap 
function as
\begin{align}
Q(O,\varrho) &= \lam_3 - t_1\; (\lam_3 - \lam_1) - t_2\; (\lam_3 - \lam_2) \;.
\label{M3:defQandtk}\end{align}
Then, we need to calculate
\begin{align}
\mcP_{\bm{\lam}}(t_1,t_2) &= \frac{1}{Z(\bm{o})} 
   \left\{ {\prod_{k=1}^2} \int\frac{\dd s_k}{2\pi}\right\}\; Z(\bm{s})\; , \\
Z(\bm{s})&= \int\frac{\dd^2\tau}{4\pi^2}
   \frac{1}{\prod_{j=1}^3 \det[\bm{C}(\lam_j)]}\; ,
\end{align}
where the matrix $\bm{C}(\lam)$ is defined as 
\begin{equation}
\bm{C}(\lam) = \left(\begin{array}{cc} 
c_1(\lam) & \tau/2 \\
-\tau^*/2 & c_2(\lam) \end{array}\right)\; , 
\end{equation}
with $c_j(\lam)= 1 - is_j (\lam - t_j)$. As in Ref.~\cite{AG16}, let us start 
with the integral
\begin{equation}
\int\frac{\dd s_1}{2\pi}\; 
\frac{1}{\prod_{j=1}^3 \det[\bm{C}(\lam_j)]},
\end{equation}
which can be calculated quite simply with the help of the residue theorem. To 
this end, we note that each determinant gives rise to one single pole at the 
position
\begin{align}
\det\, \bm{C}(\lam) = 0 \ &\Leftrightarrow\ 1 - is_1(\lam - t_1) 
                    = \frac{-|\tau|^2}{4\, c_2(\lam)}\nonumber\\
&\Leftrightarrow\ s_1 = \frac{i}{t_1 - \lam }
    \left[ \,1 +\frac{|\tau|^2}{4\, c_2(\lam)}\, \right]\; .
\end{align}
Even though $c_2(\lam)$ is complex, this does not affect the sign of the 
imaginary part of the pole. The pole lies on the upper (lower) half plane for 
$\lam < t_1$ ($\lam > t_1$). Since we may close the integration path either
over the upper or the lower half plane, the integral is non-zero only if the
two poles are lying on opposite sides of the real axis. This implies that there
are only two cases of a non-zero result:
\begin{align}
&\text{a)} \quad 0 < \lam_1 < t_1 < \lam_2 < \lam_3\; , \nonumber\\
&\text{b)} \quad 0 < \lam_1 < \lam_2 < t_1 < \lam_3\; . \nonumber
\end{align}
In both cases, the evaluation of all integrals is a rather tedious 
exercise, the description of which can be found in App.~\ref{appB1} 
and App.~\ref{appB2}, with the respective results given in the 
Eqs.~(\ref{M3D:A:res}) and~(\ref{M3D:B:res}). These can be combined into one 
single expression:
\begin{align}
\mcP_{\bm{\lam}}(t_1,t_2) &= \frac{1}{h}\begin{cases} 
   F(\lam_1,\lam_3,\lam_2) - F(\lam_1,\lam_2,\lam_3) &: \text{a)}\\
   F(\lam_1,\lam_3,\lam_2) - F(\lam_3,\lam_2,\lam_1) &: \text{b)}
   \end{cases} \; ,
\label{CentralRes}\end{align} 
where
\begin{align}
F(a, b, c) &= |t_1 - a + t_2 - b | +  |t_2 - c |\; , \notag\\
\text{and}\quad 
h &= (\lam_{1} - \lam_{2})(\lam_{1}-\lam_{3})(\lam_{2}-\lam_{3}) \; .
\label{M3:defFandh}\end{align}
Equation~(\ref{CentralRes}) provides the central result of this work. While it
is probably difficult to determine this JPD directly, it allows to calculate
the overlap distribution $P_\varrho(q)$ in closed form; see Sec.~\ref{MQ}, 
below.

\begin{figure}
  \includegraphics[scale = .38]{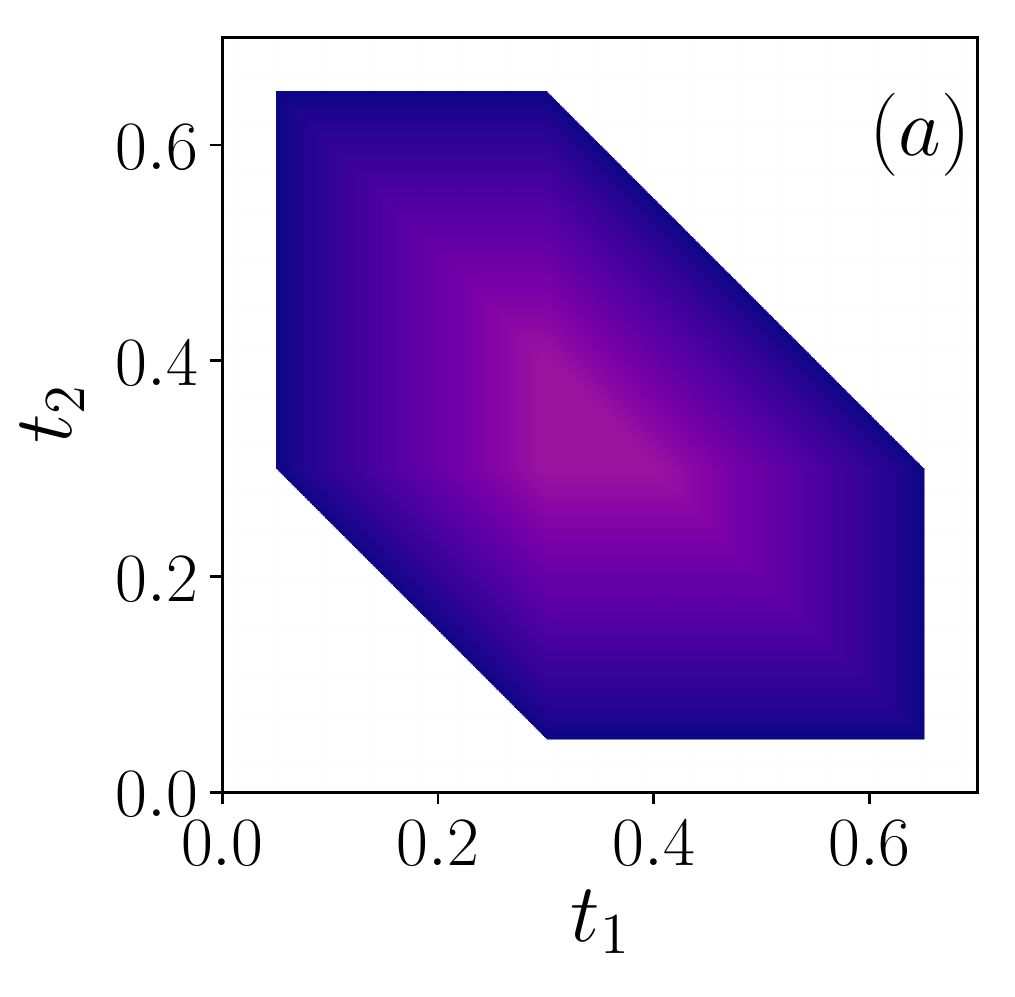}
  \includegraphics[scale = .38]{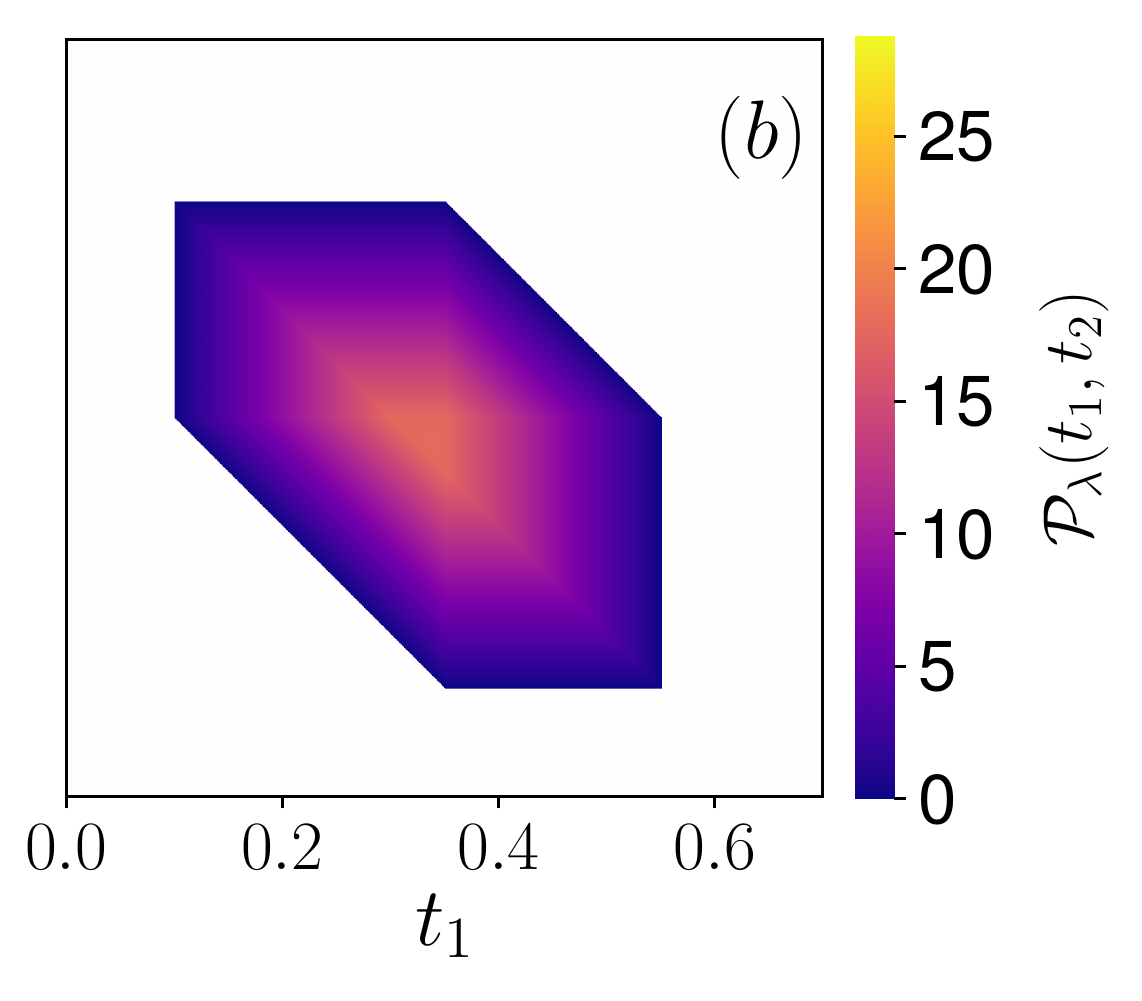}\\
  \includegraphics[scale = .38]{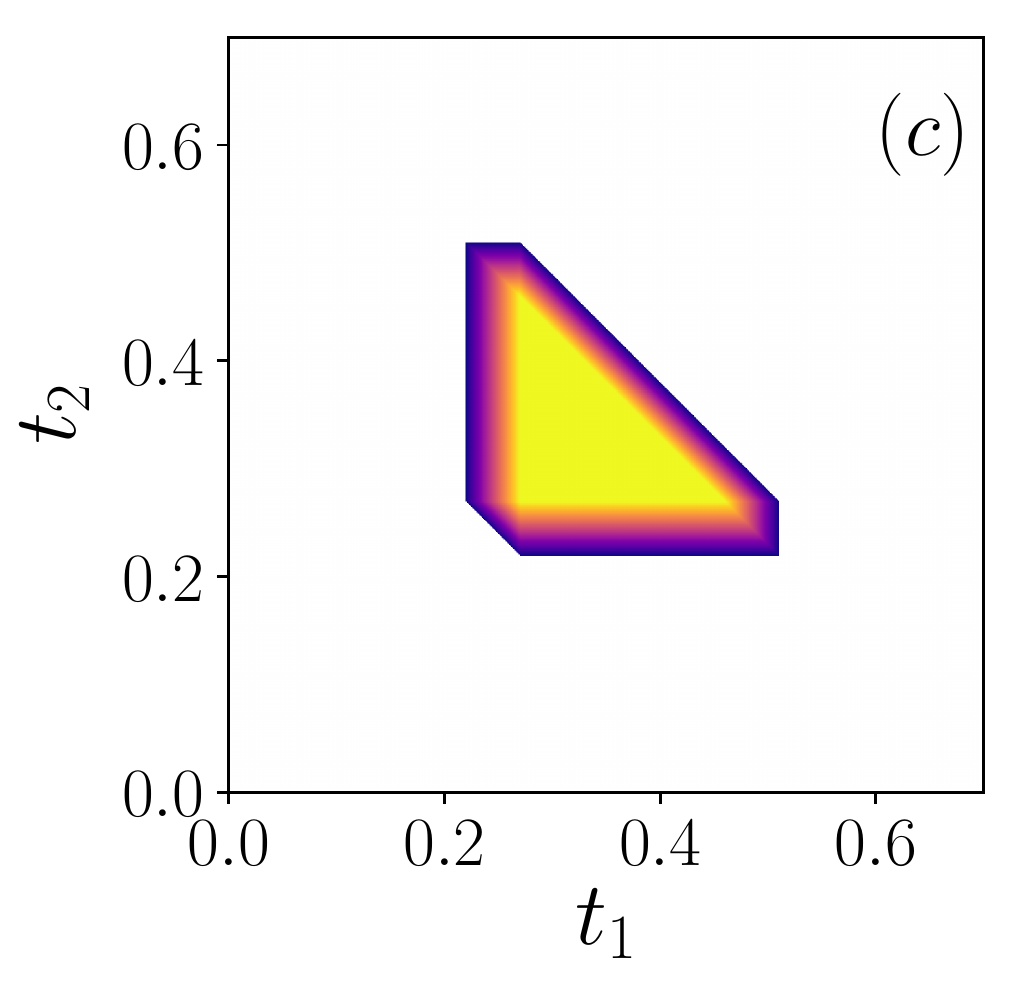}
  \includegraphics[scale = .38]{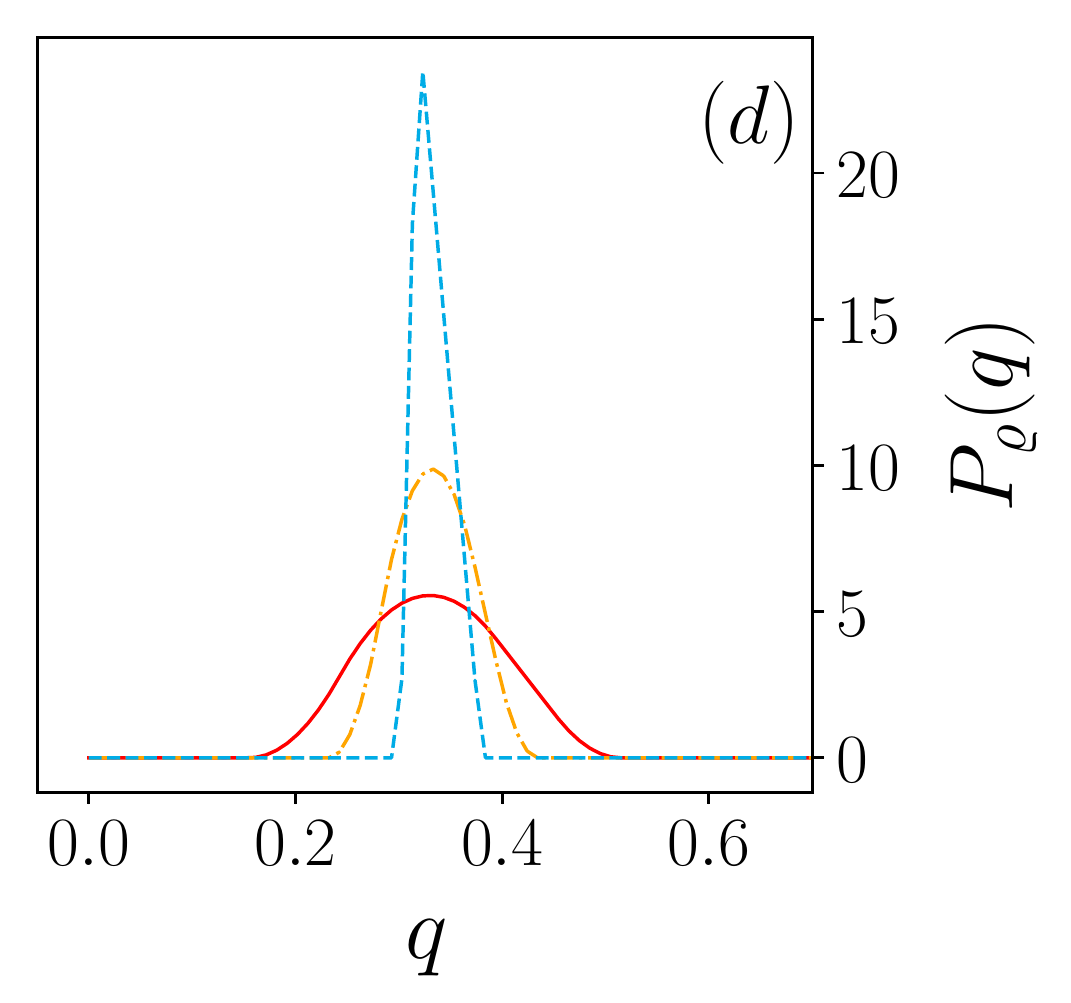}
\caption{(Color online) Joint probability density $\mcP_{\bm\lam}(t_1,t_2)$ for 
  different sets of eigenvalues: (a) $\lambda_1 = 0.05, \; \lambda_2 = 0.3 $, 
  (b) $\lambda_1 = 0.1, \; \lambda_2 = 0.35$, and (c) 
  $\lambda_1 = 0.22, \; \lambda_2 = 0.27$. Panel (d) shows the corresponding
  overlap distribution $P_{\varrho}(q)$. Red solid line (a), orange dot-dashed
  line (b), and blue dashed line (c).}
\label{plambda_plots}\end{figure}

In Fig.~\ref{plambda_plots}, panels (a), (b) and (c), we show 
$\mcP_{\bm\lam}(t_1,t_2)$ for different sets of eigenvalues. The color coding
is such that low (high) probability densities are represented by dark (bright)
colors, while areas where the probability density is equal to zero are left
white. This allows to identify that region in the $(t_1,t_2)$ plane, where the
probability density, $\mcP_{\bm\lam}(t_1,t_2)$, is larger than zero. The region 
of non-zero probability density may be characterized as the intersection of two 
simple geometric sets, the rectangular region 
$\lambda_1 < t_1, t_2 < \lambda_3$ and the 
infinite stripe $\lambda_1 + \lambda_2 < t_1 + t_2 < \lambda_2 + \lambda_3$.
The surface defined by $\mcP_{\bm\lam}(t_1,t_2)$ consists of seven 
planar triangles, stitched together to form a continuous polyhedron with an 
open hexagonal base and a horizontal top central triangle. When two eigenvalues 
become equal, but different from the third, the polyhedron reduces to a right 
triangular prism, and when all three eigenvalues become equal, the JPD becomes 
a two-dimensional delta distribution, centered at $t_1 = t_2 = 1/3$. The 
overlap distribution $P_{\varrho}(q)$, shown in panel~(d), will be discussed
separately in Sec.~\ref{MQ}.

\subsection{\label{MQ} Overlap distribution function for a qutrit}

The JPD for partial overlaps $t_1$ and $t_2$, discussed in the previous
section is not directly measurable, but it allows to calculate the overlap 
probability density.
\begin{align}
P_{\varrho}(q) &= \langle \delta (q - Q(O, \varrho)) \rangle \nonumber\\
&= \int_{0}^{1} {\rm {d}} t_1 \int_{0}^{1} {\rm {d}} t_2\; 
   \mathcal{P}_{\lambda}(t_1, t_2)\; \delta (q - Q(O, \varrho) )\; .
\end{align}
The variable transformation, 
\[ t_2 \to q' = \lambda_3 - (\lambda_3 - \lambda_1)\; t_1 
      - (\lambda_3 - \lambda_2)\; t_2\; , \]
allows us to evaluate the $t_2$-integral eliminating the delta-function:
\begin{align}
P_{\varrho}(q) &= \frac{1}{\lam_{3} - \lam_{2}} \int_{0}^{1} {\rm {d}} t_1 
   \int_{\lam_2 -(\lam_3 - \lam_1)t_1}^{\lam_3 -(\lam_3 - \lam_1)t_1} {\rm {d}} 
   q' \nonumber\\
&\times\mathcal{P}_{\bm\lambda}
 \left(t_1, \frac{\lam_3 - q' - (\lam_3 - \lam_1)t_1}{\lambda_3 - \lambda_2}
 \right) \delta (q - q') \; .
\end{align}
Since $q'$ is limited to the interval 
\[ \big (\, 
     \lam_2 - (\lam_3 - \lam_1) t_1 \, ,\, \lam_3 - (\lam_3 - \lam_1)t_1\, 
   \big )\; , \]
the $q'$ integral contributes to the overlap probability density, only if $q$
is lying in the same interval. That leads to the following additional 
restrictions for $t_1$:
\begin{equation}
t_1 < \frac{\lambda_3 - q}{\lambda_3 - \lambda_1}\; , \quad\text{and}\quad 
t_1 > \frac{\lambda_2 - q}{\lambda_3 - \lambda_1}\; .
\end{equation}
Therefore, we obtain 
\begin{align}
P_\varrho(q)= \int_{t_1^{\min}}^{t_1^{\max}} {\rm d} t_1\; 
   \mathcal{P}_{\bm\lam}\left(t_1, 
      \frac{\lam_3 - q - (\lam_3 - \lam_1)t_1}{\lam_3 - \lam_2}\right)\; ,
\label{MO:Prhoint}\end{align}
where the new integration bounds are given by
\begin{align}
t_1^{\min} &= \max\left(0, \frac{\lam_2 - q}{\lam_3 - \lam_1} \right)\; ,
% \\
\quad
t_1^{\max} = \min\left(1, \frac{\lam_3 - q}{\lam_3 - \lam_1} \right)\; .
\end{align}
Since $\mathcal{P}_{\bm\lam}(t_1,\alpha + \beta\, t_1)$ is a piecewise linear 
function, the integration is straightforward for any particular case. As a 
result, $P_\varrho(q)$ becomes piecewise quadratic. In 
Fig.~\ref{plambda_plots}, panel (d),  $P_{\varrho}(q)$ is plotted for the
combination of eigenvalues considered in the first three panels of 
Fig.~\ref{plambda_plots}.

\subsection{\label{MR} Range of the overlap distribution}

The range of the overlap probability density, $P_\varrho(q)$ is limited to a 
finite interval: $q_{\rm min} < q < q_{\rm max}$, with limits depending on
the eigenvalues of $\varrho$. Here, we will show that it is typically enough
to know these limits, in order to estimate the eigenvalues of $\varrho$. 

\begin{figure}
\centering
\includegraphics[scale = .4]{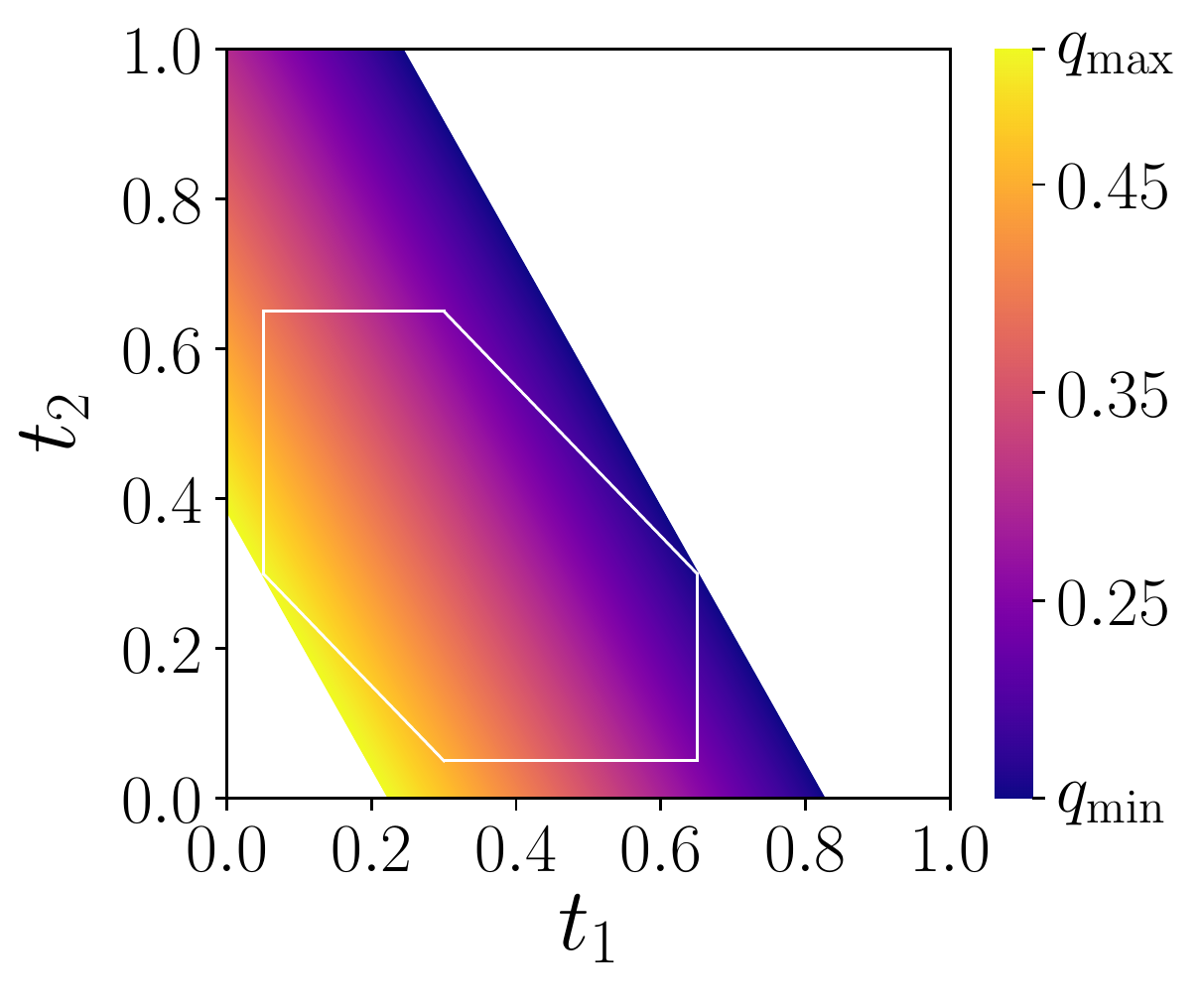}
\caption{(Color online) The overlap $q(t_1,t_{2})$ as a color coded density 
plot, as defined in Eq.~(\ref{MO:equipot}), for $\lambda_1 = 0.05$ and 
$\lambda_2 = 0.3$. In this case, $q_{\rm min} = 0.155$ and 
$q_{\rm max} = 0.515$.}
\label{qdist}\end{figure}

Based on Eq.~(\ref{M3:defQandtk}), we may write the overlap $Q(O,\varrho)$ as a 
function of the partial overlaps, $t_1,t_2$ defined in Eq.~(\ref{overlapFunc}):
\begin{align}
q(t_1,t_2)= \lam_3 - t_1\; (\lam_3 - \lam_1) - t_2\; (\lam_3 - \lam_2) \; .
\label{MO:equipot}\end{align}
In Fig. \ref{qdist}, we show a density plot of this function, where we chose
for the eigenvalues the case (a) from Fig.~\ref{plambda_plots}. The white lines 
indicate the limits of the range of $\mathcal{P}_{\bm\lam}(t_1, t_2)$. It can be seen that equipotential 
lines are straight lines, with the overlap value $q$ increasing towards the 
origin of the $(t_1,t_2)$-plane. On the basis of the discussion 
below Eq.~(\ref{M3:defFandh}), one readily verifies that the maximum value 
$q_{\rm max}$ is reached at the point $(t_1,t_2) = (\lambda_1, \lambda_2)$, 
whereas the minimum value is reached at $(t_1,t_2) = (\lambda_3, \lambda_2)$.
Consequently, the maximum (minimum) value of $q$ is given by
\begin{align}
q_{\rm max} = \lambda_1^2 + \lambda_2^2 + \lambda_3^2 
\quad\text{and}\quad
q_{\rm min} = \lambda_2^2 + 2\lambda_1\lambda_3 \; ,
\label{qmaxqmin}\end{align}
respectively. 
The upper limit $q_{\rm max}$ is of course just the purity of the mixed state
$\varrho$, which can be obtained by choosing the identity for the unitary
transformation in $Q(O,\varrho)$. To the best of our knowledge, the expression
for the lower limit is a new result.

\section{\label{S}Stability of the eigenvalue estimation}

Let us consider the task of estimating the eigenvalues of a mixed quantum state 
$\varrho$ from the statistics of overlaps. For that 
purpose, we assume that an experiment is realized, where overlaps are measured 
for a large but finite sample of random unitary transformations. The result of
the experiment may be represented in the form of a histogram, which should
approximate the true overlap distribution $P_\varrho(q)$. Since the sample is
finite, the histogram will show deviations from the true distribution, and this 
will lead to uncertainties in the estimation of the eigenvalues.

A particularly simple procedure consists in using the experimental data to 
estimate $q_{\rm min}$ and $q_{\rm max}$, with uncertainties 
$\Delta q_{\rm min}$ and $\Delta q_{\rm max}$, respectively. In the case of
qutrits, the knowledge of these two quantities is sufficient to determine the 
desired eigenvalues of the mixed quantum state under investigation. In what 
follows, we assume normally distributed errors, to compute the propagated 
uncertainties on these eigenvalues.

Resolving Eq.~(\ref{qmaxqmin}) for $\lambda_1$ and $\lambda_2$, taking into 
account that $\lambda_1 +\lambda_2 + \lambda_3 = 1$, we find
\begin{align}
\lambda_2 &= \frac{1 - \sqrt{3(q_{\rm max} + q_{\rm min} ) - 2}}{3},
\label{S:lam2}\\
\lambda_1 &= \frac{1 - \lambda_2 - \sqrt{q_{\rm max} - q_{\rm min}}}{2} \; .
\end{align}
To work out the error propagation for $\lambda_1$ and $\lambda_2$, let us 
introduce the auxiliary parameters,
\begin{equation}
r= q_{\rm max} + q_{\rm min}\; , \qquad
s= q_{\rm max} - q_{\rm min}\; ,
\end{equation}
which have both the same statistical error
\begin{equation}
\Delta^2 = \Delta q_{\rm min}^2 +\Delta q_{\rm max}^2 \; .
\end{equation}
Then, we obtain
\begin{align}
(\Delta\lambda_2)^2 = \left( \frac{\partial \lambda_2}{\partial r}\right)^2\;
\Delta^2 = \frac{r^2}{4\, (3r-2)}\; \Delta^2 \; .
\end{align}
Similarly,
\begin{align}
(\Delta\lambda_1)^2 &= \left( \frac{\partial \lambda_1}{\partial r}\right)^2\;
\Delta^2  + \left( \frac{\partial \lambda_1}{\partial s}\right)^2\; 
\Delta^2 \notag\\
&
= \left( \frac{r^2}{3r-2} + \frac{1}{s}\right) \frac{\Delta^2}{16} \; .
\end{align}
Finally,
\begin{equation}\label{uncertantyLambs}
\frac{(\Delta\lambda_2)}{\Delta} = \frac{r}{2\, \sqrt{3r-2}} \; , \quad
\frac{(\Delta\lambda_1)}{\Delta} = \frac{1}{4}\,
\sqrt{\frac{r^2}{3r-2} + \frac{1}{s}} \; .
\end{equation}
These two expressions show that the uncertainty on the estimated eigenvalues 
can be kept small, as long as $r \ne 2/3$ and $s\ne 0$. According to 
Eq.~(\ref{S:lam2}), the first condition implies that $\lambda_2 = 1/3$, 
$\lambda_1 + \lambda_3 = 2/3$; remember the ordering of the eigenvalues:
$\lambda_1 < \lambda_2 < \lambda_3$. The second condition implies that
$q_{\rm max} = q_{\rm min}$ which means that all eigenvalues must be equal.
This situation is a limiting case of the first condition.

\section{\label{W}\boldmath Overlap function for Wigner $D$-matrices}

In this section, we assume that the unitary transformations to be applied to 
the system are limited to $SO(3)$, the subgroup of rotations in $3D$ coordinate 
space. Examples of such systems are qutrits build from photon 
pairs~\cite{Bur03}.

A natural parametrization for the subgroup $SO(3)$, is provided by the Wigner 
$D$-matrices for angular momentum quantum number $j = 1$.
\begin{equation}
D = \left( \begin{array}{ccc} 
      \frac{1 + \cos\beta}{2} e^{-i(\alpha + \gamma)} & 
      \frac{- e^{-i\alpha}}{\sqrt{2}} \sin\beta  & 
      \frac{1 - \cos\beta}{2} e^{-i(\alpha - \gamma)} \\ 
  \frac{e^{-i\gamma}}{\sqrt{2}}\sin\beta & 
  \cos\beta & 
  \frac{-e^{i\gamma}}{\sqrt{2}} \sin\beta \\ 
      \frac{1 - \cos\beta}{2} e^{i(\alpha - \gamma)} & 
      \frac{e^{i\alpha}}{\sqrt{2}} \sin\beta & 
      \frac{1 + \cos\beta}{2} e^{i(\alpha + \gamma)} \end{array} \right).
\end{equation}
Thus, we are interested in the distribution of 
\begin{equation}
 Q(D,\varrho) = {\rm Tr}\big [\, D\, \varrho\, D^\dagger\; \varrho\, \big ] 
\; . 
\end{equation}
Here, we can no longer assume that some transformation $D$ will diagonalize 
$\varrho$. As a consequence, the distribution of overlaps will depend not only 
on the eigenvalues of $\varrho$, but also on its eigenstates. To make this 
dependence clear, we assume that $\varrho$ is diagonalized by the unitary 
matrix $U$: $\varrho = U\, \underline{\lambda}\, U^\dagger$. Then, we may write
\begin{align}
Q(D,\varrho) &= Q(D,U,\underline{\lambda}) 
= \text{Tr}\left[ D\, U\, \underline{\lambda}\, U^{\dagger}\, D^{\dagger}\, 
U \underline{\lambda}\, U^{\dagger}\right]\notag\\
&= \text{Tr}\left[ \tilde D\,\underline{\lambda}\,  \tilde D^\dagger\, 
\underline{\lambda}\right], \quad \tilde D = U^{\dagger}\, D\, U \; .
\label{W:QintermsTiD}\end{align}
Varying the angles $\alpha,\beta,\gamma$, moves $\tilde D$ along a 
three-dimensional orbit within the group $SU(3)$, where the 
orbit itself is determined by the eigenvector matrix $U$.

Our aim is the estimation of the eigenvalues of $\varrho$ from the 
overlap distribution $P_\varrho(q)$, under a random (or systematic) sampling 
over the Wigner $D$-matrices. This is now much more complicated, since
$P_\varrho(q)$ now also depends on the eigenvectors of $\varrho$, collected as
column vectors in $U$. 

For instance, in the previous section we showed that it is sufficient to 
determine $q_{\rm min}$ and $q_{\rm max}$ in order to estimate the eigenvalues, 
since there we got two equations for two unknowns. However, here we have 
many additional unknowns for the eigenvectors. Therefore, on should expect that
the determination of $q_{\rm min}$ and $q_{\rm max}$ will no longer be 
sufficient. In what follows, we use again the JPD of partial overlaps, 
$\mathcal{P}_\varrho(t_1,t_2)$ together with the full overlap distribution 
$P_\varrho(q)$, to analyze this problem.

\subsection{\label{WR} Density matrices with randomly chosen eigenvectors}

\begin{figure}
  \includegraphics[scale = .4]{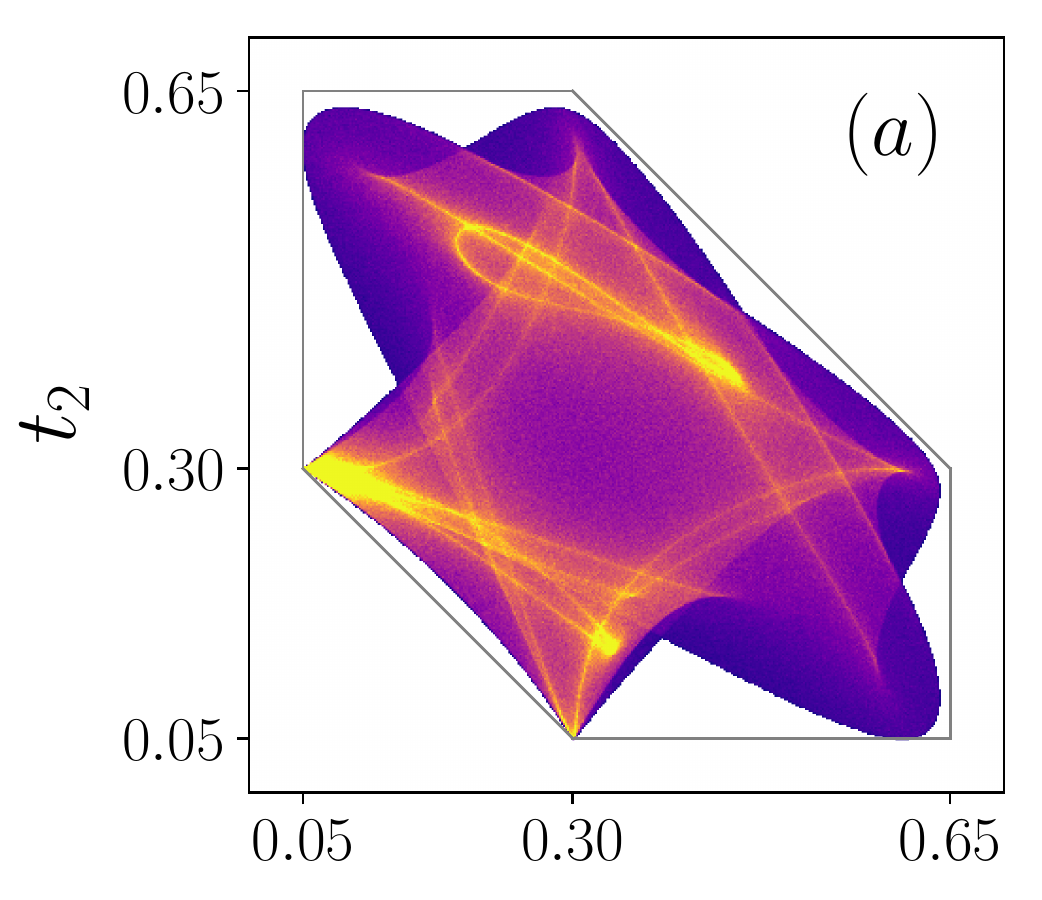}
  \includegraphics[scale = .4]{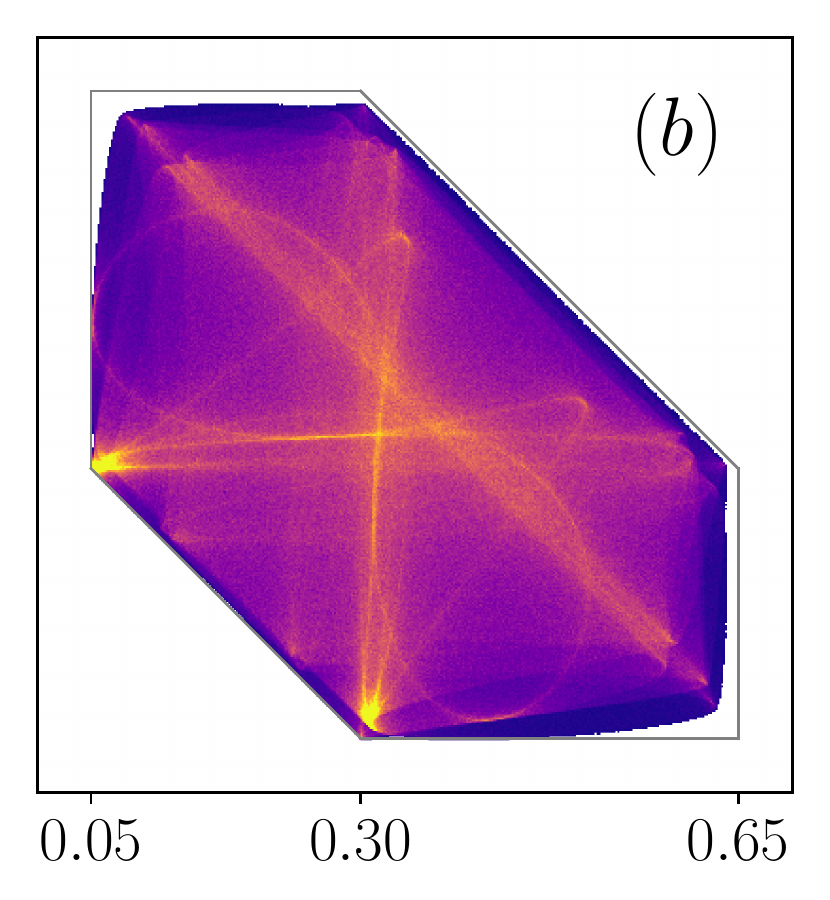}\\
  \includegraphics[scale = .4]{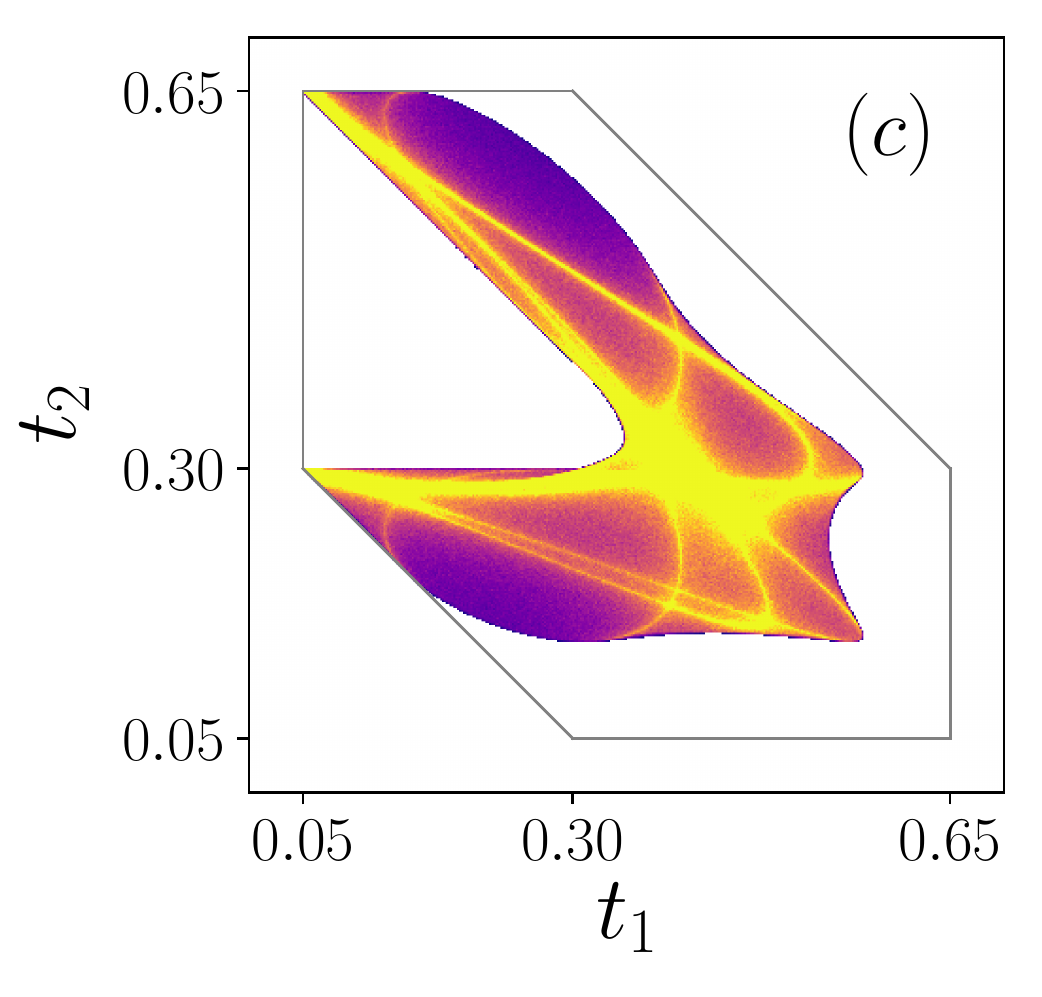}
  \includegraphics[scale = .4]{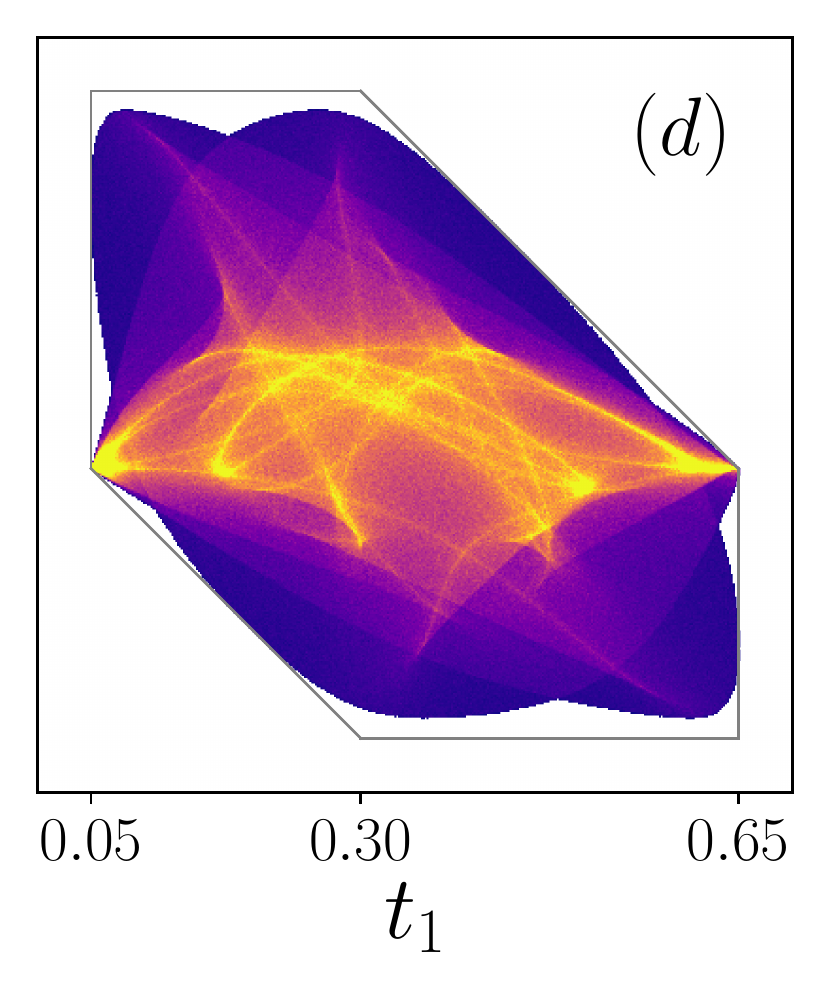}
\caption{(Color online) Density of points $(t_1, t_2)$, Eq. (\ref{SO3deftk}), 
for different random unitary matrices; panels (a), (b), (c) and (d). The color 
coding reflects the density of points, such that regions with a higher (lower) 
density are plotted in yellow (purple). In the white regions no hits were 
found.} 
\label{moreOverlap}\end{figure}

In this section, we consider density matrices with fixed eigenvalues 
$\lambda_1 = 0.05, \lambda_2 = 0.3$ (the same eigenvalues as in 
Figs.~\ref{plambda_plots} and~\ref{qdist}), but choose their eigenvector matrix
$U$ at random from the full group $SU(3)$. We then compute a large sample 
(10 million elements) of pairs of partial overlaps, $(t_{1},t_{2})$ according
to the expression 
\begin{equation}
t_k = \lambda_{1}\; |\,\tilde{D}_{1k}\,|^2
    + \lambda_{2}\; |\,\tilde{D}_{2k}\,|^2 
    + \lambda_{3}\; |\,\tilde{D}_{3k}\,|^2 \; ,
\label{SO3deftk}\end{equation}
obtained from Eq.~(\ref{W:QintermsTiD}). Here, each pair $(t_{1}, t_{2})$ is
obtained from a Wigner $D$-matrix, with angles $\alpha,\beta,\gamma$ chosen 
from uniform distributions. 

In Fig.~\ref{moreOverlap} we plot an approximation to 
$\mathcal{P}_\varrho(t_1,t_2)$, in the form of a density plot for the sample 
of partial overlaps mentioned above. For each of the four panels, a different 
random matrix $U$ of eigenvectors has been chosen, and all of them for the same 
combination of eigenvalues $\lam_{1} = 0.05$ and $\lambda_2 = 0.3$. These plots 
are obtained by color coding the number of points $(t_1,t_2)$ hitting the area
of each pixel on the graph (dark regions, mean a low density of points; bright 
regions a high density; regions without points were left white). The $SU(3)$ 
limits for the support of $\mathcal{P}_\varrho(t_1,t_2)$, obtained in 
Sec.~\ref{M}, are plotted by thin solid lines. 

In all cases, it is possible to reach $q_{\rm max}$ 
at $(t_1,t_2) = (\lambda_1,\lambda_2)$. This is because the identity is an 
element of $SO(3)$ also. By contrast, it may not be possible to reach 
$q_{\rm min}$ at $(t_1,t_2) = (\lambda_3,\lambda_2)$. A very clear example for 
that is shown in panel (c). By contrast, panel (d) shows a case, where it seems 
that $q_{\rm min}$ is reached.

\begin{figure}
\includegraphics[width=0.45\textwidth,height= 0.3\textwidth]
   {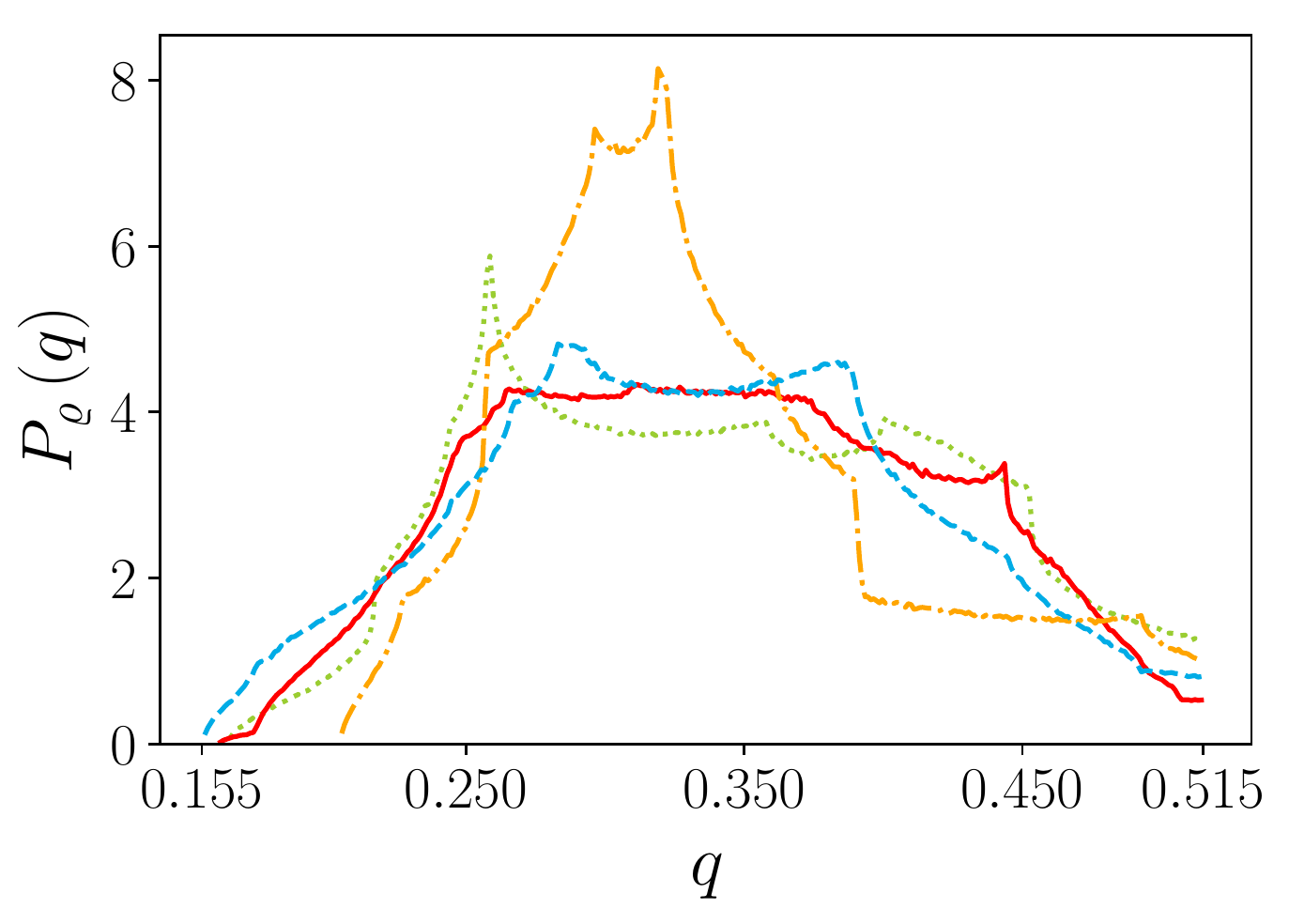}
\caption{(Color online) Overlap distribution $P_\varrho(q)$ for the cases, 
corresponding to the four panels in Fig.~\ref{moreOverlap}. Green dotted line
[panel (a)], red solid line [panel (b)], orange dot-dashed line [panel (c)],
and blue dashed line [panel (d)].}
\label{pdistunitarias}\end{figure}	

Similarly, in Fig.~\ref{pdistunitarias}, we plot an approximation to the total 
overlap distribution $P_\varrho(q)$, in the form of a histogram over the total 
overlaps $q$, computed according to Eq.~(\ref{M3:defQandtk}). The histograms 
are calculated from the four data sets shown in Fig.~\ref{moreOverlap}. The
different distributions posses a number of particularities, which may serve as 
a fingerprint for the density matrix under investigation. This means that using 
the overlap distribution, it is easy to distinguish different density matrices. 
However, so far we are not aware of any method to solve the inverse problem of
determining eigenvalues and orientation of the eigenvectors, given the 
overlap distribution.
  
As mentioned before, in all cases, the overlap distribution reaches the 
maximal value $q_{\rm max}$, which is simply the purity of the density matrix 
in question. However, depending on the orientation of the eigenvectors, the 
minimum value $q_{\rm min}$ may or may not be reached; {\rm c.f.} discussion of
Fig.~\ref{moreOverlap}.
  
In Sec.~\ref{MR}, we discussed the error propagation in the estimation of the 
eigenvalues of $\varrho$, based on measured values for $q_{\rm min}$ and 
$q_{\rm max}$. Based on this discussion, we find that it may be possible to 
obtain at least approximate estimates for the eigenvalues of $\varrho$, if 
the minimum value for $Q(D,\varrho)$ over the Wigner-$D$ matrices is 
sufficiently close to the absolute minimum, $q_{\rm min}$.

\begin{figure}
\includegraphics[scale = .55]{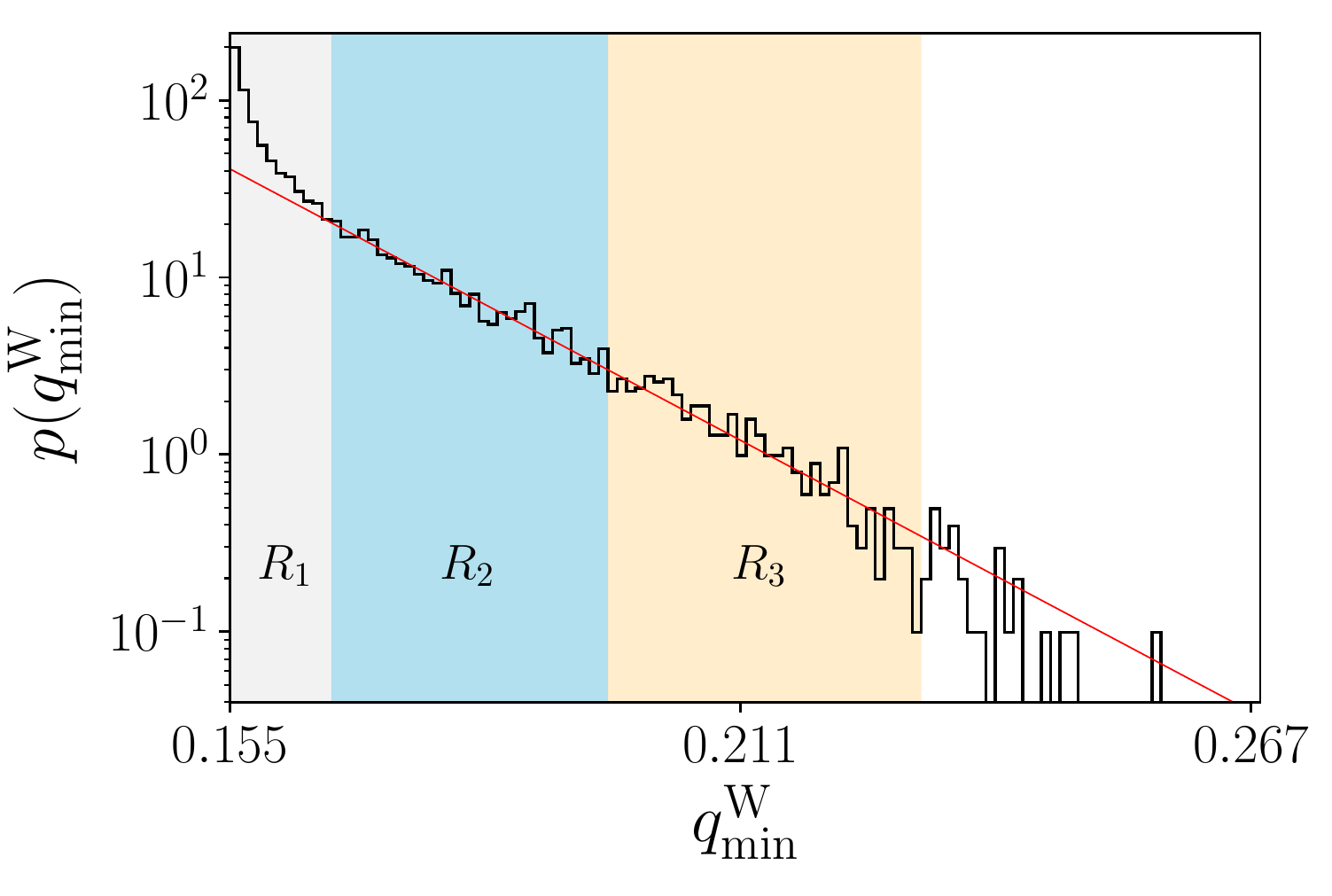}
\caption{(Color online) Probability distribution for $q_{\rm min}$  over an 
ensemble of unitary matrices. The red line is fitted to the histogram's tail, 
see main text for more details.}
\label{distqminqmaxt}\end{figure}

For that reason we study the distribution of minimum values, 
\begin{align}
q^{\rm W}_{\rm min} = \min_D Q(D,\varrho)\; ,
\end{align}
in Fig.~\ref{distqminqmaxt}. For that purpose, we chose $10^4$ eigenvector 
matrices $U$ at random, searching for each one the minimum of $Q(D,\varrho)$ 
over $10^5$ randomly chosen Wigner matrices. Of course, this provides only a
numerical approximation to the true distribution of minimum values 
$q^{\rm W}_{\rm min}$. The resulting distribution, again for $\lambda_1= 0.05$ 
and $\lambda_2= 0.3$, is shown in Fig.~\ref{distqminqmaxt}. We checked that a 
duplication of the sample of Wigner $D$-matrices would yield an 
indistinguishable result.  

The minimum values $\min_D Q(D,\varrho)$ are restricted to the interval 
$[q_{\rm min}, q_{\rm wc}]$, where the latter is calculated in the following
Sec.~\ref{WW}. Both quantities only depend on the eigenvalues of the mixed 
state $\varrho$ to be analyzed. The latter, $q_{\rm wc}$, is the largest 
possible minimum value $q^{\rm W}_{\rm min}$, which depends on the eigenvector
matrix $U$. In Fig.~\ref{distqminqmaxt}, it can be seen that for the majority
of eigenvector matrices, $q^{\rm W}_{\rm min}$ is rather close to 
$q_{\rm min} = 0.155$ and rather far away from $q_{\rm wc} = 0.267$, which
means that with high probability, we may obtain useful estimates for the 
eigenvalues of $\varrho$, with reasonably small uncertainties. To illustrate 
this fact, we highlight the regions $R_1$, $R_2$ and $R_3$, with the following
significance: In region $R_1$ we have $68.3\%$  
of the counts, in $R_1 + R_2$ it is $95.4\%$,   
and if we combine all the regions,  $R_1 + R_2 + R_3$ the count is $99.7\%$.
These percentage values are taken from the familiar $68$-$95$-$99.7$ rule of 
statistical significance tests.

\subsection{\label{WW} Maximin overlap}

In game theory, ``maximin'' is a term used to describe the strategy which
maximizes one's own minimum gain. Similarly, we consider here the maximum
of the minimum overlap, where the minimum is taken over the Wigner 
$D$ matrices, and the maximum subsequently over all possible eigenvector 
matrices of $\varrho$. Thus
\begin{align}
q_{\rm wc} = \max_{U \in SU(3)}\; \min_D\; Q(D,\varrho)
   = \max_{U \in SU(3)}\; q_{\rm min}^{\rm W} \; . 
\end{align}
Depending on the eigenvectors of $\varrho$, $q_{\rm min}^{\rm W}$ may be 
considerably larger than the absolute minimum value $q_{\rm min}$, calculated 
in Sec.~\ref{MR}, as has been shown in Fig.~\ref{distqminqmaxt}.

In the present section, we search for those eigenvector matrices $U$ that lead 
to the largest minimum value $q_{\rm wc}$. We conjecture that we may restrict
our search to the set of permutation matrices:
\begin{align}
U_{(1,3)} &= \begin{pmatrix} 0 & 0 & 1\\ 0 & -1 & 0\\ 1 & 0 & 0\end{pmatrix}
  \; , \qquad
U_{(1,2)} = \begin{pmatrix} 0 & 1 & 0\\ 1 & 0 & 0\\ 0 & 0 & -1\end{pmatrix}
\; , \notag\\
U_{(2,3)} &= \begin{pmatrix} -1 & 0 & 0\\ 0 & 0 & 1\\ 0 & 1 & 0\end{pmatrix}
\; ,
\end{align}
all of which belong to $SU(3)$, due to the diagonal element, set to minus one.
From these three permutation matrices, only the first one belongs to the 
class of Wigner-$D$ matrices. We thus expect, that for a density matrix with
such eigenvectors, the value of $q_{\rm min}$ can be reached. For the other
two cases, the minimum values $q_{\rm min}^{\rm W}$ are larger, and we 
conjecture that the largest of the two will be $q_{\rm wc}$.

\subsubsection{Permutation $(1,3)$}

We simply calculate the general expression for $\tilde D$, according to 
Eq.~(\ref{W:QintermsTiD}), and obtain $t_1$, $t_2$ according to 
Eq.~(\ref{SO3deftk}):  %TG for $U_{(1,3)}$ equal to the identity:
\begin{align}
t_1 
&=  \lambda_2 \frac{\sin^{2}\beta}{2} + (\lambda_1 + \lambda_3) \frac{1 + \cos^2\beta}{4} 
+ (\lambda_1 - \lambda_3) \frac{ \cos\beta}{2},\\
t_2
&=  \lambda_2 \cos^{2}\beta + (\lambda_1 + \lambda_3)  \frac{\sin^{2}\beta}{2}.
\end{align}
This describes a single one-dimensional line in the space of partial overlaps,
parametrized by $\beta$, one of the Wigner angles. This line connects the 
points, where the overlap becomes maximal ($q_{\rm max}$) and where it becomes
minimal ($q_{\rm min}$), as shown in Fig. \ref{overlapDist} (a) (solid line).
Hence, in this case $q_{\rm min}^{\rm W} = q_{\rm min}$.

\subsubsection{ Permutation  $(1, 2)$}

\begin{figure}
	\includegraphics[scale = .34]{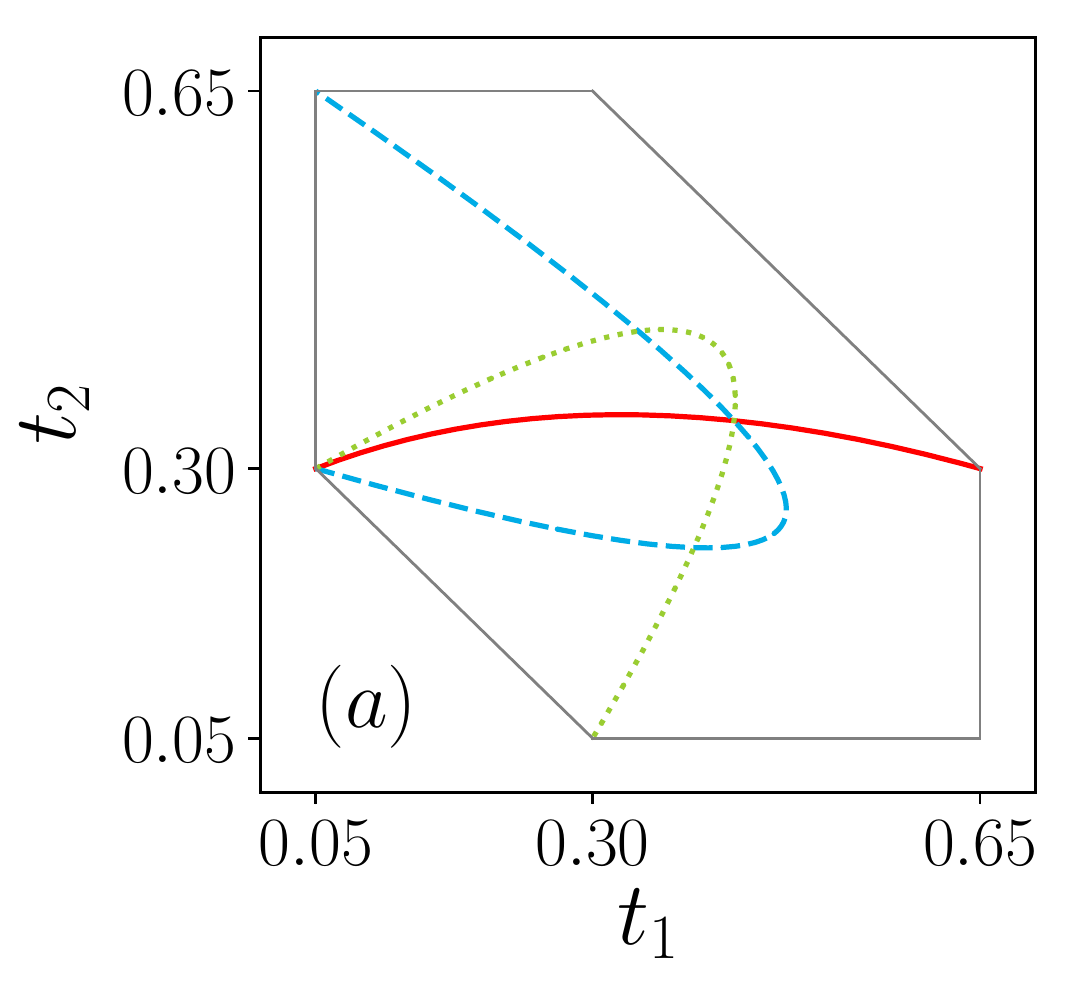}
	\includegraphics[scale = .34]{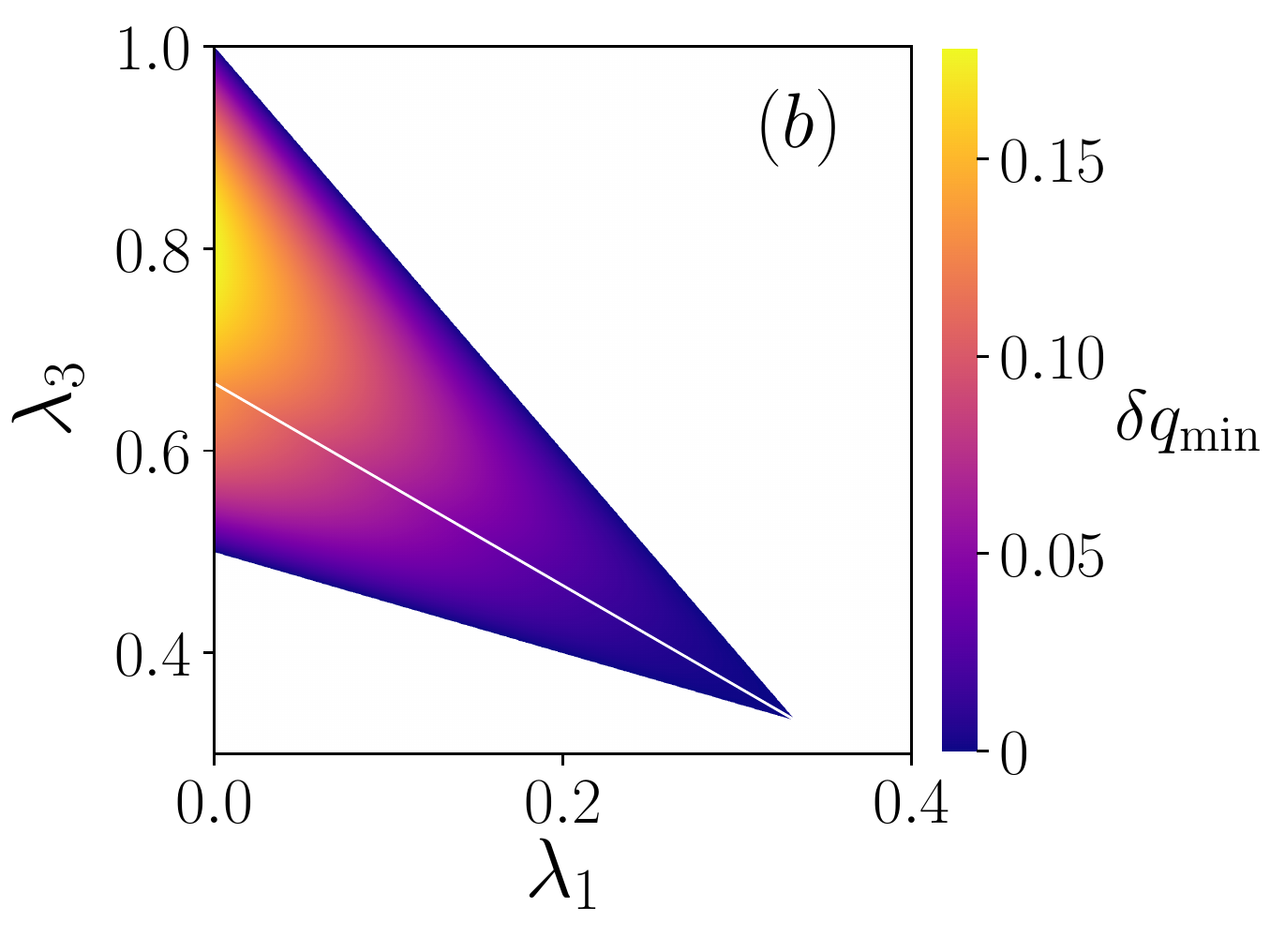}
	\caption{(Color online) (a) Set of points $(t_1, {t_2})$  which can be 
 reached by $SO(3)$ transformations on $\varrho$, for different eigenvector
 matrices $U$. Permutation $(1,3)$ (solid line), permutation $(1,2)$ (dashed 
 line), and permutation $(2,3)$ (dotted line). (b) The largest difference
 $\delta q_{\rm min} = q_{\rm wc} - q_{\rm min}$, as a function of the 
 eigenvalues of $\varrho$. }
\label{overlapDist} \end{figure}

Again, we calculate the general expression for $\tilde D$
according to Eq.~(\ref{W:QintermsTiD}) and subsequently $t_1, t_2$ according to
Eq.~(\ref{SO3deftk})
\begin{align}
t_1 &= -\, \frac{1 - 3\lambda_1}{2}\; \cos^2\beta
+ \frac{1 - \lambda_1}{2}  \notag\\
t_2 &= \frac{1 - 3\lambda_1}{4}\; \cos^2\beta
+ \frac{1 - \lambda_1 - 2\lambda_3}{2}\; \cos\beta
+ \frac{1 + \lambda_1}{4} \; . 
\end{align}
Again, we obtain a one-dimensional curve parametrized by the Wigner angle 
$\beta$; see Fig.~\ref{overlapDist}, panel~(a) (dashed line). According to
Eq.~(\ref{M3:defQandtk}), we find for the overlap,
\begin{align}
q &= \frac{(1 - 3\lambda_1)^2}{4}\; \cos^2\beta
+ \frac{(1 - \lambda_1 - 2\lambda_3)^2}{2}\; \cos\beta \notag\\
&\qquad + \frac{(1 - \lambda_1)\, (1 + 3\lambda_1)}{4}\; .
\end{align}
To find $q_{\rm min}^{\rm W}$ in this case, we have to find the minimum of $q$
as a function of $\beta$. Thus, we use the derivative to find the extremal 
values
\begin{align}
&\frac{-1}{\sin\beta}\, \frac{{\rm d} q}{{\rm d}\beta} 
= \frac{(1 - 3\lambda_1)^2}{4}\; 
2\cos\beta\,
- \frac{(1 - \lambda_1 - 2\lambda_3)^2}{2} = 0 \notag\\
&\Leftrightarrow\quad \cos\beta = -\, 
\frac{(1 - \lambda_1 - 2\lambda_3)^2}{(1 - 3\lambda_1)^2} \; .
\end{align}
Of course $\beta = 0$ and $\pi$ are also extremal points, but they correspond
to $q$ reaching its maximal value. Inserting this result into the equation
for $q$, we find
\begin{align}
q^{(1,2)}_{\rm min} &= -\, 
\frac{(1 - \lambda_1 - 2\lambda_3)^4}{4\, (1 - 3\lambda_1)^2}
+ \frac{(1 - \lambda_1) (1 + 3\lambda_1)}{4}\; .
\label{qmin12}
\end{align}

\subsubsection{Permutation $(2,3)$} 

Following the same steps as before, we find expressions for $t_1$, $t_2$ and 
$q^{(2,3)}_{\rm min} $. Explicitly, we obtain
\begin{align}
q^{(2,3)}_{\rm min} &= -\, 
\frac{(1 - \lambda_3 - 2\lambda_1)^4}{4\, (1 - 3\lambda_3)^2}
+ \frac{(1 - \lambda_3) (1 + 3\lambda_3)}{4}\; ,
\label{qmin23}
\end{align}
and the resulting curve in $(t_1, t_2)$-space can be seen in 
Fig.~\ref{overlapDist} (a) as a dotted line. 

As mentioned before, we conjecture that all other choices of eigenvector 
matrices $U$ lead to smaller values for the minimal overlaps over the Wigner
$D$-matrices, {\rm i.e.,}
\begin{equation}
 Q(D, \varrho) \le \max\left(\, q_{\rm min}^{(1,2)}\, , \, 
      q_{\rm min}^{(2,3)}\right) \; ,
\end{equation}
which is supported by the numerical test given in Fig.~\ref{distqminqmaxt}.
Therefore, provided the conjecture holds, we may write
\begin{equation} 
q_{\rm wc} = \max\left( q^{(1, 2)}_{\rm min}\, , \, q^{(2, 3)}_{\rm min} \right)
\; .
\label{worstcase:qmin}
\end{equation}  
For $\lambda_1 = 0.05$ and $\lambda_2 = 0.3$, $q_{\rm w} = 0.267$, which is
taken as the upper tick-mark for the horizontal axis in 
Fig.~\ref{distqminqmaxt}. 

In Fig.~\ref{overlapDist}(b), we plot 
$\delta q_{\rm min} = q_{\rm wc} -  q_{\rm min}$ as a function of the 
eigenvalues $\lambda_1$ and $\lambda_3$, where the value of 
$\delta q_{\rm min}$ is color coded as indicated in the legend of the figure.
The white line from $(0,2/3)  \quad\mbox{to}\quad  (1/3, 1/3)$ marks the 
border between the regiones where $q_{\rm wc} = q^{(1, 2)}_{\rm min}$ (above)
and $q_{\rm wc} = q^{(2, 3)}_{\rm min}$ (below). 

\section{\label{refC} Conclusions}

In this paper we considered the distribution of overlaps between a mixed state
and its image under random unitary transformations. For general unitary
transformations from $SU(N)$ we showed that the overlap distribution can be
computed, in principle, following a method introduced in Ref.~\cite{AG16}.
The solution is based on the calculation of the joint probability distribution
(JPD) of partial overlaps, for which we could obtain a surprisingly simple 
analytical expression in the qutrit, {\rm i.e.,} $SU(3)$, case. Based on this 
result, we obtained a closed expression for the distribution of overlaps, and 
we solved the inverse problem of estimating the eigenvalues from the overlap 
distribution and computed the corresponding error estimates.

In the second part of the paper, we assumed that the random transformations are 
restricted to the subgroup $SO(3)$, in the form of Wigner $D$-matrices. Then,
the overlap distribution also depends on the eigenvectors. Choosing some random
examples, we find JPD's for the partial overlaps, with many particular 
features, which are reflected in the full overlap distribution, also. These may 
serve as fingerprints for particular mixed quantum states, allowing to verify 
their identity.

We then focused on the finite range of the overlap distribution, and showed 
that the $SO(3)$ limits often come quite close to the absolute $SU(3)$ limits, 
which means that while not exact, eigenvalue estimates based on transformations 
limited to $SO(3)$ may be useful, provided the error is sufficiently small.
Finally, we conjecture that Eq.~(\ref{worstcase:qmin}) gives the largest 
possible difference between the minimal overlap calculated from $SO(3)$ and
that calculated from $SU(3)$.

\begin{acknowledgments}
It is a pleasure to thank A. Klimov for drawing our attention to this problem 
and for very fruitful discussions.
\end{acknowledgments}

\appendix
\section{\label{G3} Limiting cases for a qutrit \boldmath ($N = 3$)}

According to the general expression for the overlap function, 
Eq.~(\ref{G:QforNm1}), we obtain for the qutrit case 
\begin{align}
Q(O,\varrho) &= \lam_3 - (\lam_3- \lam_1)\; t_1 - (\lam_3 - \lam_2)\; t_2\; ,
\end{align}
where $t_1$ and $t_2$ are given by Eq.~(\ref{overlapFunc}). Even without the 
general solution, we can still solve the following limiting cases. Eventually,
we will need the distribution function $\mathcal{P}_{NK}(t)$ of the sum of 
absolute-value squares of $K$ elements of column vectors of $O$ in $N=3$ 
dimensions. These functions have been calculated in Ref.~\cite{AG16}.

\subsection{\boldmath $\lam_1 = \lam_2 = \lam_3 = 1/3$}
For this case we simply have
\begin{equation}
Q(O,\varrho) = \frac{1}{3},\quad 
   P_{\varrho}(q) = \langle \delta(q - 1/3)\rangle\; .
\end{equation}

\subsection{\boldmath $\lam_1 = 0, \lam_2 = \lam_3 = 1/2$}
For this case we have,
\begin{align}
Q(O,\varrho)
&= \frac{1}{2}(1- t_1) \; , \notag\\
\mbox{with}\quad t_1 &= \frac{|O_{21}|^2 +  |O_{31}|^2}{2}
 = \frac{1 - |O_{11}|^2}{2} \; ,
\end{align}
since the matrix $O$ is unitary. Therefore,
\begin{align}
Q(O,\varrho) &= \frac{1+ |O_{11}|^2}{4}\; .
\end{align}
With this result, the probability density becomes
\begin{align}
P_{\varrho}(q) &= \langle\; \delta\left( q - Q(O,\varrho)\right) \;\rangle 
   =  \lla \delta\left( q - \frac{1+ |O_{11}|^2}{4}\right) \;\rra\nonumber\\
&= \int_{0}^{1} {\rm {d}}t \;\mathcal{P}_{31}(t)\;  
      \delta\left( q - \frac{1 + t}{4}\right)\nonumber\\
&=  4 \int_{1/4}^{1/2} {\rm {d}}q' \;\mathcal{P}_{31}(4q' -1)\;  
      \delta\left( q -q'\right)\; ,
\end{align}
and finally
\begin{equation}
P_{\varrho}(q) = 4 \mathcal{P}_{31}(4q-1)\quad\text{for}\quad 
   1/4 \leq q \leq 1/2
\end{equation}
and zero otherwise.

\subsection{\boldmath $\lam_1 = \lam_2 = 0, \lam_3 = 1$}

We have
\begin{align}
Q(O,\varrho) &= 1 - t_1 -t_2, \nonumber\\
&= 1 - \left(|O_{31}|^2 + |O_{32}|^2\right) = |O_{33}|^2,
\end{align}
with $t_1 = |O_{31}|^2$, $t_2 = |O_{32}|^2$, and $|O_{31}|^2 + |O_{32}|^2 + |O_{33}|^2=1$. Hence,
\begin{equation}
Q(O,\varrho) = |O_{33}|^2,\quad P_{\varrho}(q) = \mathcal{P}_{31}(q)\; .
\end{equation}

\section{Details of the JPD calculation for \boldmath $N=3$ \label{appB}}

Here, we will provide some details on the evaluation of the integrals, which
finally lead to Eq.~(\ref{CentralRes}) for the JPD of the partial overlaps in 
the qutrit case. The starting point is the following integral
\begin{equation}
\int\frac{\dd s_1}{2\pi}\; \frac{1}{\prod_{j=1}^3 \det[\bm{C}(\lam_j)]}.
\end{equation}
The general method is shown here for the two orderings that give non-trivial results using formulas in Appendix \ref{appC} when needed.

\subsection{\boldmath 
   Case a) $\quad 0 < \lam_1 < t_1 < \lam_2 < \lam_3$ \label{appB1}}
Here the three poles are given by
\begin{equation}
z_j =  \frac{i}{t_1 - \lam_j }\left[1 + \frac{|\tau|^2}{4\, c_2(\lam_j)}\, \right], \quad j=1,2,3,
\end{equation}
such that $z_1$ lies on the upper half plane, while $z_2$ and $z_3$ are lying on the lower one. Thus, we close the integration over the upper half plane, which results in a positively oriented closed curve around the pole at $z_1$.

Using the formula form Appendix \ref{apps1}, the integral over $s_1$ can be written as
\begin{equation}
\int\frac{\dd s_1}{2\pi}\; 
\frac{1}{\prod_{j=1}^3 \det[\bm{C}(\lam_j)]} =\frac{-(\lam_1-t_1)c_2(\lam_1)}{(a_{12}+b_{12}|\tau|^2)(a_{13}+b_{13}|\tau|^2)},
\end{equation}
where we have defined
\begin{align}
a_{jk}&\equiv (\lam_j-\lam_k)c_2(\lam_j)c_2(\lam_k),\\
b_{jk}&\equiv\frac{1}{4}[(\lam_j-t_1)c_2(\lam_j)-(\lam_k-t_1)c_2(\lam_k)].
\end{align}

Next, we perform the integral over $\tau$ on the result using the formula in Appendix \ref{apptau},
\begin{align}
-(\lam_1-t_1)&c_2(\lam_1)\int \frac{\dd \tau^2}{4\pi^2}
\frac{1}{(a_{12}+b_{12}|\tau|^2)(a_{13}+b_{13}|\tau|^2)}\nonumber\\
&=\frac{-(\lam_1-t_1)c_2(\lam_1)}{4\pi}\frac{\ln(a_{12}b_{13})-\ln(a_{13}b_{12})}{a_{12}b_{13}-a_{13}b_{12}}.
\end{align}

Now, we want to perform the final integral in $s_2$, the dependence on this variable is hidden in $a_{jk}$ and $b_{jk}$ through $c_2(\lam_k)$. Let us make this dependence explicit by the replacements
\begin{align}
a_{12}b_{13}-a_{13}b_{12}=& \frac{1}{4}(t_1-\lam_1)(\lam_1-\lam_2)(\lam_1-\lam_3)\nonumber\\
&\times(\lam_2-\lam_3)s_2^2[1+is_2(t_2-\lam_1)],
\end{align}
where
\begin{equation}
\frac{a_{12}b_{13}}{a_{13}b_{12}}=\frac{1+is_2(t_2-\lam_2)}{1+is_2(t_2-\lam_3)}\cdot
\frac{1+is_2(t_1-\lam_1+t_2-\lam_3)}{1+is_2(t_1-\lam_1+t_2-\lam_2)},
\end{equation}
simplifying, we obtain
\begin{align}
&\int \frac{\dd s_2}{2\pi} \dots = \frac{1}{2\pi^2(\lam_1-\lam_2)(\lam_1-\lam_3)(\lam_2-\lam_3)}\nonumber\\
&\times\int\frac{\dd s_2}{s_2^2}
\ln\left[\frac{1+is_2(t_2-\lam_2)}{1+is_2(t_2-\lam_3)}\cdot
\frac{1+is_2(t_1-\lam_1+t_2-\lam_3)}{1+is_2(t_1-\lam_1+t_2-\lam_2)}\right].
\end{align}
The integral is performed in Appendix \ref{apps2} and the result is
\begin{align}
\int\frac{\dd s_2}{2\pi}& \dots = \frac{|t_2 -\lambda_2|  - |t_2 -\lambda_3|}
{2\pi(\lam_1-\lam_2)(\lam_1-\lam_3)(\lam_2-\lam_3)}\nonumber\\
&+\frac{|t_1 +t_2 -\lambda_1 -\lambda_3| - |t_1+ t_2 -\lambda_1 -\lambda_2|}{2\pi(\lam_1-\lam_2)(\lam_1-\lam_3)(\lam_2-\lam_3)}.
\end{align}

Then, in order to obtain $\mcP_{\lam}(t_1,t_2)$ we must calculate $Z(\bm{0})$, given by $Z(\bm{s})$ in Eq. \eqref{zetas} with $s_j=0$. It is then given by the following integral that can be solved straightforward
\begin{equation}
Z(\bm{0})=\int\frac{\dd \tau^2}{4\pi^2}\frac{1}{\left(1+\frac{|\tau|^2}{4}\right)^3}=\frac{1}{2\pi}.
\end{equation}

Thus, finally for ordering \textbf{(a)}
\begin{align}
\mcP_{\lam}(t_1,t_2)&=\frac{|t_1 +t_2 -\lambda_1 -\lambda_3| - |t_1+ t_2 -\lambda_1 -\lambda_2|} {(\lam_1-\lam_2)(\lam_1-\lam_3)(\lam_2-\lam_3)}\nonumber\\
&+\frac{|t_2 -\lambda_2|  - |t_2 -\lambda_3|}{(\lam_1-\lam_2)(\lam_1-\lam_3)(\lam_2-\lam_3)}.
\label{M3D:A:res}
\end{align}

\subsection{\boldmath
   Case b) $\quad 0 < \lam_1 < \lam_2 < t_1 < \lam_3$ \label{appB2}}
Here, there are two poles in the upper half plane given by $z_1$ and $z_2$. We can calculate the first integral on $s_1$ with the residue theorem again but now with two poles, i.e.,
\begin{align}
&\int\frac{\dd s_1}{2\pi}\; 
\frac{1}{\prod_{j=1}^3 \det[\bm{C}(\lam_j)]} \nonumber\\
&=\left. \frac{i}{\det \bm{C}(\lam_2)\det \bm{C}(\lam_3)}
\right|_{s_1 = z_1}\; 
{\rm Res}\left( \frac{1}{\det[\bm{C}(\lam_1)]},z_1\right)\nonumber\\
&+\left. \frac{i}{\det \bm{C}(\lam_1)\det \bm{C}(\lam_3)}
\right|_{s_1 = z_2}\; 
{\rm Res}\left( \frac{1}{\det[\bm{C}(\lam_2)]},z_2\right).
\end{align}
The first integral is exactly the one we calculated in case \textbf{(a)}. The second integral has a similar form and we can use the same integration method. The second residue is
\begin{equation}
{\rm Res}\left( \frac{1}{\det[\bm{C}(\lam_2)]},z_2\right)=\frac{i}{(\lam_2-t_1)c_2(\lam_2)},
\end{equation}
and using the formula from Appendix \ref{apps1}, we can calculate $\left.\det\bm{C}(\lam_k)\right|_{s_1=z_2}$ for $k=1,3$. With these equations, we can obtain that the new part of the integral in $s_1$, given by the residue in $z_2$ is
\begin{align}
&\left. \frac{i}{\det \bm{C}(\lam_1)\det \bm{C}(\lam_3)}\right|_{s_1 = z_2}
{\rm Res}\left( \frac{1}{\det[\bm{C}(\lam_2)]},z_2\right)\nonumber\\
&\qquad\qquad=\frac{-(\lam_2-t_1)c_2(\lam_2)}{(a_{21}+b_{21}|\tau|^2)(a_{23}+b_{23}|\tau|^2)}.
\end{align}

The integral over $\tau$ is solved using the formula from Appendix \ref{apptau} as
\begin{align}
-(\lam_2-t_1)&c_2(\lam_2)\int \frac{\dd \tau^2}{4\pi^2}\frac{1}{(a_{21}+b_{21}|\tau|^2)(a_{23}+b_{23}|\tau|^2)}\nonumber\\
&=\frac{-(\lam_2-t_1)c_2(\lam_2)}{4\pi}\frac{\ln(a_{21}b_{23})-\ln(a_{23}b_{21})}{a_{21}b_{23}-a_{23}b_{21}}.
\end{align}

Replacing the definitions of $a_{jk}$ and $b_{jk}$ for $c_2(\lam_j)$ and $c_2(\lam_k)$ and then to $s_2$, we obtain
\begin{align}
&a_{21}b_{23}-a_{23}b_{21}= \frac{1}{4}(t_1-\lam_2)(\lam_2-\lam_1)(\lam_2-\lam_1)(\lam_1-\lam_3)\nonumber\\
&\hspace{27mm}\times s_2^2[1+is_2(t_2-\lam_2)],\\
&\frac{a_{21}b_{23}}{a_{23}b_{21}}=\frac{1+is_2(t_2-\lam_1)}{1+is_2(t_2-\lam_3)}\cdot
\frac{1+is_2(t_1-\lam_2+t_2-\lam_3)}{1+is_2(t_1-\lam_2+t_2-\lam_1)},
\end{align}
simplifying, we obtain
\begin{align}
&\int\frac{\dd s_2}{2\pi} \dots = \frac{1}{2\pi^2(\lam_2-\lam_3)(\lam_2-\lam_1)(\lam_1-\lam_3)}\nonumber\\
&\times\int\frac{\dd s_2}{s_2^2}
\ln\left[\frac{1+is_2(t_2-\lam_1)}{1+is_2(t_2-\lam_3)}\cdot
\frac{1+is_2(t_1-\lam_2+t_2-\lam_3)}{1+is_2(t_1-\lam_2+t_2-\lam_1)}\right].
\end{align}
This integral in $s_2$ is given by the formula in Appendix \ref{apps2}, therefore
\begin{align}
\int&\frac{\dd s_2}{2\pi} \dots =\frac{|t_2-\lam_1|-|t_2-\lam_3|}
{2\pi(\lam_2-\lam_3)(\lam_2-\lam_1)(\lam_1-\lam_3)}\nonumber\\
&+\frac{|t_1+t_2-\lam_2-\lam_3|-|t_1+t_2-\lam_1-\lam_2|}
{2\pi(\lam_2-\lam_3)(\lam_2-\lam_1)(\lam_1-\lam_3)}.
\end{align}
Using the calculated value of $Z(\bm{0})=(2\pi)^{-1}$ and the value of the integral of part \textbf{(a)}, we obtain
\begin{align}
\mcP_{\lam}(t_1,t_2)=&\frac{|t_1 +t_2 -\lambda_1 -\lambda_3| - |t_1+ t_2 -\lambda_2 -\lambda_3|} {(\lam_1-\lam_2)(\lam_1-\lam_3)(\lam_2-\lam_3)}\nonumber\\
&+\frac{|t_2 -\lambda_2|  - |t_2 -\lambda_1|} {(\lam_1-\lam_2)(\lam_1-\lam_3)(\lam_2-\lam_3)}.
\label{M3D:B:res}
\end{align}

\section{Integration formulas \label{appC}}

\subsection{Integrals over \boldmath $s_1$ \label{apps1}} 
The following integral
\begin{align}
\int\frac{\dd s_1}{2\pi}\frac{1}{\prod_{j=1}^3 \det[\bm{C}(\lam_j)]}&
= \left. \frac{i}{\det \bm{C}(\lam_2) \det \bm{C}(\lam_3)}\right|_{s_1 = z_1}\nonumber\\
&\times{\rm Res}\left( \frac{1}{\det[\bm{C}(\lam_1)]}, z_1\right),
\end{align}
where the residue is given by
\begin{equation}
{\rm Res}\Big (\frac{1}{\det[\bm{C}(\lam_j)]}, z_j\Big )=\frac{i}{(\lam_j - t_1)\, c_2(\lam_j)},
\end{equation}
in this case with $j=1$. The evaluation of $\det\bm{C}(\lam_k)$ at $s_1=z_j$ is given by
\begin{equation}
\left.\det\bm{C}(\lam_j)\right|_{s_1=z_j}=a'_{jk}+b'_{jk}|\tau|^2,
\end{equation}
with
\begin{align}
a'_{jk}&\equiv\frac{\lam_j-\lam_k}{\lam_j-t_1}c_2(\lam_k),\\
b'_{jk}&\equiv\frac{1}{4}\left(1-\frac{\lam_k-t_1}{\lam_j-t_1}\frac{c_2(\lam_k)}{c_2(\lam_j)}\right).
\end{align}

After simplifying, the integral over $s_1$ can be written as
\begin{equation}
\int\frac{\dd s_1}{2\pi}\; 
\frac{1}{\prod_{j=1}^3 \det[\bm{C}(\lam_j)]} =\frac{-(\lam_1-t_1)c_2(\lam_1)}{(a_{12}+b_{12}|\tau|^2)(a_{13}+b_{13}|\tau|^2)},
\end{equation}
where we have defined
\begin{align}
a_{jk}&\equiv (\lam_j-\lam_k)c_2(\lam_j)c_2(\lam_k),\\
b_{jk}&\equiv\frac{1}{4}[(\lam_j-t_1)c_2(\lam_j)-(\lam_k-t_1)c_2(\lam_k)].
\end{align}
\subsection{\boldmath $\tau$-integral \label{apptau}}
This integral is
\begin{equation}
\int \frac{\dd \tau^2}{4\pi^2} \frac{1}{(a_1+b_1|\tau|^2)(a_2+b_2|\tau|^2)}
\end{equation}

We write $\tau$ in polar coordinates as $\tau=r \text{e}^{i\theta}$ to integrate
\begin{align}
\int_0^{2\pi}& \frac{\dd \theta}{4\pi^2}
\int_0^\infty \frac{r\dd r}{(a_1+b_1r^2)(a_2+b_2r^2)}\nonumber\\
&=\frac{1}{2\pi}
\left.\frac{\ln(a_1+b_1r^2)-\ln(a_2+b_2r^2)}{2(a_2b_1-a_1b_2)}\right|_0^\infty\nonumber\\
&=\frac{1}{4\pi}\frac{\ln(a_1b_2)-\ln(a_2b_1)}{a_1b_2-a_2b_1}.
\end{align}

\subsection{Integrals over \boldmath $s_2$ \label{apps2}}
The $s_2$ integral is given by
\begin{align}
\int_{-\infty}^\infty \frac{\dd x}{x^2} & \ln
\left[\frac{1+iq_1x}{1+iq_2x} \cdot \frac{1+iq_3x}{1+iq_4x}\right] \notag\\
= &\; \pi\left( |q_1|  - |q_2| + |q_3| - |q_4|\right).
\end{align}

\end{document}